\documentclass[showpacs,preprintnumbers,nofootinbib,amsmath,amssymb]{revtex4}
 \usepackage{graphicx}
\usepackage{dcolumn}
\usepackage{enumitem}
\usepackage{bm}
\begin{document}
\title{Neutrino-nucleus cross-sections at supernova neutrino energies}
\author{S. Chauhan, M. Sajjad Athar\footnote{Corresponding author: sajathar@gmail.com} and S. K. Singh}
\affiliation{Department of Physics, Aligarh Muslim University, Aligarh - 202 002, India}

\begin{abstract}
The inclusive neutrino/antineutrino-induced charged and neutral current reaction cross-sections in $^{12}C$, $^{16}O$, $^{40}Ar$, 
$^{56}Fe$ and $^{208}Pb$ in the energy region of supernova neutrinos/antineutrinos are studied. The calculations are performed in the local density approximation
(LDA) 
 taking into account the effects due to Pauli blocking, Fermi motion and the renormalization of weak transition 
strengths in the nuclear medium. The effect of Coulomb distortion of the lepton produced in the charged current reactions 
has also been included. The numerical results for the energy dependence of the cross-section $\sigma(E)$ as well as the flux averaged cross-section and 
event rates for the charged lepton production in the case of some supernova neutrino/antineutrino fluxes recently discussed in the literature have 
been presented. We have also given the flux-averaged angular and energy distributions of the charged leptons corresponding to these fluxes.
\end{abstract}
\maketitle
\section{Introduction}
Supernova explosion is a phenomenon which occurs in the late phase of stellar evolution. In this explosion, most 
of the gravitational energy released in a core collapse is carried by the neutrinos. Such neutrino bursts carry 
about $\approx 2.5 \times 10^{53}$ ergs of energy in a very short period of time \cite{Raffelt}. It is 
considered that these neutrinos provide valuable information about the proto-neutron star core, its equation 
of state, core collapse and supernova explosion mechanism and help to have a better understanding 
of the supernova physics~\cite{Raffelt, Haxton}. After the observation of supernova neutrinos from SN1987A in Kamiokande, IMB and BAKSAN 
\cite{Hirata:1987hu, Bionta:1987qt}, 
the feasibility of detecting such events in the future is now given serious consideration. 
 For example, the experiments like SuperKamiokande (SK)~\cite{Bays:2012iy}, Large Volume Detector (LVD)~\cite{Agafonova:2007hn}, Antarctic Muon And
Neutrino Detector Array (AMANDA)~\cite{Amanda}, Boron solar neutrino experiment (BOREXino)~\cite{Borexino}, Observatory for Multiflavour Neutrino 
Interactions from Supernovae (OMNIS)~\cite{Smith:2001ns}, Lead Astronomical Neutrino Detector (LAND)~\cite{Hargrove:1996zv}, 
 Helium and Lead Observatory (HALO)~\cite{HALO}, Imaging Cosmic And Rare Underground Signals (ICARUS)~\cite{Icarus}, etc., are in various stages of 
 operation while Sudbury Neutrino Observatory (SNO+)~\cite{SNO+}, Hyper-Kamiokande (Hyper-K)~\cite{Hyperk}
 experiments are being developed, and experiments like Deep Underground Neutrino Experiment (DUNE)~\cite{Strait:2016mof} and 
Jiangmen Underground Neutrino Observatory (JUNO)~\cite{Djurcic:2015vqa, An:2015jdp} are planned to study the physics related to 
supernova neutrinos in the near future.
A list of the present and future experiments having sensitivity to the supernova neutrinos/antineutrinos is given in Table \ref{tab:expt}~\cite{Scholberg:2012id}. These
experiments are planned to use detector with various nuclei as the target material. This makes the knowledge of neutrino nucleus cross-section of low 
energy neutrino/antineutrino scattering from the medium and heavy nuclear targets as an important aspect of the study of supernova neutrino detection. 
Another aspect of considerable importance in the study of supernova physics is the knowledge of neutrino/antineutrino fluxes which are determined 
by the numerical simulations of supernova explosion of stars.

The supernova neutrino/antineutrino fluxes are determined from the numerical simulations of core-collapse supernova explosion of a star and depend 
on the initial properties of the collapsing star like its mass,
 density and equation of state as well as on various physical processes controlling the explosion like initial prompt burst of neutrinos following 
 neutronization, accretion and cooling in
 late phases as well as the neutrino transport in the dense star matter~\cite{Livermore, Mezzacappa:2005ju, Janka:2006fh, Dasgupta:2008zzb}. 
 The neutrino/antineutrino fluxes are found to be sensitive to the luminosities $L_\nu$ of various neutrino/antineutrino flavors which are believed to be equal 
for all the six flavors of neutrinos/antineutrinos $\nu_e$, $\bar\nu_e$ and $\nu_x$~($x=\mu,~\tau$) due to the assumption of equipartition 
of total
 available energy amongst various flavors. Some
 recent calculations have also been done assuming luminosities for $\nu_x$ which are different from the luminosities of $\nu_e(\bar\nu_e)$ and 
 varying them in
 the range of $0.5L_\nu<L_x<2L_\nu$ and keeping the luminosities of $\nu_e$ and $\bar\nu_e$  to be the 
 same~\cite{Keil:2003sw, Fischer:2008rh, Keil:2002in, Fogli:2009rd, Vaananen:2011bf, Choubey:2010up}. The simulated neutrino/antineutrino fluxes
 and mean energies of their various flavors are in general distinct from each other due to differences in their interaction with the dense star matter
 which has an excess of neutrons over protons.
 This difference leads to $\nu_e$ loosing more energy as compared to $\bar \nu_e$, which looses more energy than $\nu_x$ ($x=\mu,~\tau$ and their 
 antineutrinos) as $\nu_x$ 
 interact only through the neutral current interaction (due to higher threshold energy for charged current reactions induced by $\nu_x$), while $\nu_e$
 and $\bar \nu_e$ interact by the neutral as well as the charged current interactions. This gives a hierarchical structure of mean neutrino/antineutrino 
 energies ($E_{\nu(\bar\nu)}$) for various flavors, i.e.,
 $\langle E_{\nu_e} \rangle < \langle E_{{\bar\nu}_e} \rangle < \langle E_{\nu_x} \rangle$. Various simulations agree on this hierarchical structure of 
 mean neutrino/antineutrino energies, but differ on the
 actual values which are generally taken to be in the range of  
 $\langle E_{\nu_e} \rangle \approx 10-12 ~MeV$, $\langle E_{{\bar\nu}_e} \rangle \approx 12-15 ~MeV$ and 
 $\langle E_{\nu_x} \rangle \approx 16-25 ~MeV$~\cite{Keil:2003sw, Fischer:2008rh, Keil:2002in, 
Fogli:2009rd, Vaananen:2011bf, Choubey:2010up}. However, a lower value of  $\langle E_{\nu_e} \rangle$
has also been obtained in recent studies~\cite{Huedepohl:2009wh}
when additional medium effects are taken into account. These additional medium effects are generated by neutrino-neutrino self 
interactions~\cite{Pantaleone:1992eq, FLUX_WEB, Kneller, Duan:2010bg, Duan:2007mv} and neutrino-matter interactions when primary neutrinos of all 
flavors propagate
through a medium of very high neutrino densities and matter densities causing flavor 
conversions~\cite{Mirizzi:2006xx, Lunardini:2001pb, Takahashi:2001dc, Takahashi:2002cm}.
The nonlinear equations of motion for neutrino propagation in the presence of neutrino-neutrino and neutrino-matter interaction potential give rise 
to collective neutrino flavor 
oscillations~\cite{FLUX_WEB, Kneller, Duan:2010bg, Duan:2007mv} which along with
flavor oscillations due to MSW matter effects~\cite{Mikheev:1986gs, Wolfenstein:1977ue} contribute to the flavor conversion of neutrino affecting the 
neutrino flux spectra.
The quantitative modifications in the spectra due to these effects depend upon the 
specific value of the theoretical input parameters used in the simulations, i.e., matter density profiles, neutrino oscillation parameters specially the 
third neutrino mixing angle $\theta_{13}$, neutrino mass hierarchy, i.e.,
being normal or inverted as well as the approximations used in solving the nonlinear equations of motion. 
\begin{table}\label{tab:expt}
\begin{tabular}{|c|c|c|c|c|}\hline\hline
Experiment & Target & Mass(kT) & Location&Live Period\\\hline
LVD~\cite{Agafonova:2007hn} & $C_{n}H_{2n}$ & 1 & Italy & 1992-Present \\
BOREXino~\cite{Borexino}& $C_{n}H_{2n}$ & 0.3 & Italy & 2005-Present\\
SNO+~\cite{SNO+} & $C_{n}H_{2n}$ & 0.8 & Canada & Future \\
JUNO~\cite{Djurcic:2015vqa, An:2015jdp} & $^{12}C$ \& $^{16}O$ & 20 & China & Future\\
Super-K~\cite{Bays:2012iy}& $H_{2}O$ & 32 &  Japan & 1996-Present \\
Hyper-K~\cite{Hyperk} & $H_{2}O$ & 540 & Japan & Future \\
ICARUS~\cite{Icarus}& $^{40}Ar$ & 0.6 & Italy & 2010 - Present\\
ArgoNeuT~\cite{Anderson:2011ce} & $^{40}Ar$ & $3\times10^{-4}$ & USA & 2008 - Present \\
LAr1~\cite{Georgia} & $^{40}Ar$ & 1 &  USA & Future\\
GLADE~\cite{Georgia} & $^{40}Ar$ & 5 & USA & Future\\
LArTPC\cite{Katori:2011uq} & $^{40}Ar$ & 0.17 & USA & 2015 - Present\\
DUNE~\citep{Strait:2016mof} & $^{40}Ar$ & 68 & USA & Future \\
OMNIS~\cite{Smith:2001ns}& $^{208}Pb$ & 12 & New Mexico & Future\\
LAND~\cite{Hargrove:1996zv} & $^{208}Pb$ & 1 & Canada &Future \\
HALO~\cite{HALO} & $^{208}Pb$ & 0.076 &  Canada & Future\\\hline\hline
\end{tabular}
\caption{List of present and future neutrino detectors.}
\end{table}
Thus, the simulated neutrino flavor spectra at the surface of the star are subjected to various uncertainties
of theoretical parameters used in the simulation studies of the explosion and propagation of neutrino in the dense star matter leading to 
large variations in the predicted spectra for various flavors of neutrino/antineutrino ~\cite{Keil:2003sw, Fischer:2008rh}.

\begin{figure}
\includegraphics[height=.4\textheight,width=0.9\textwidth]{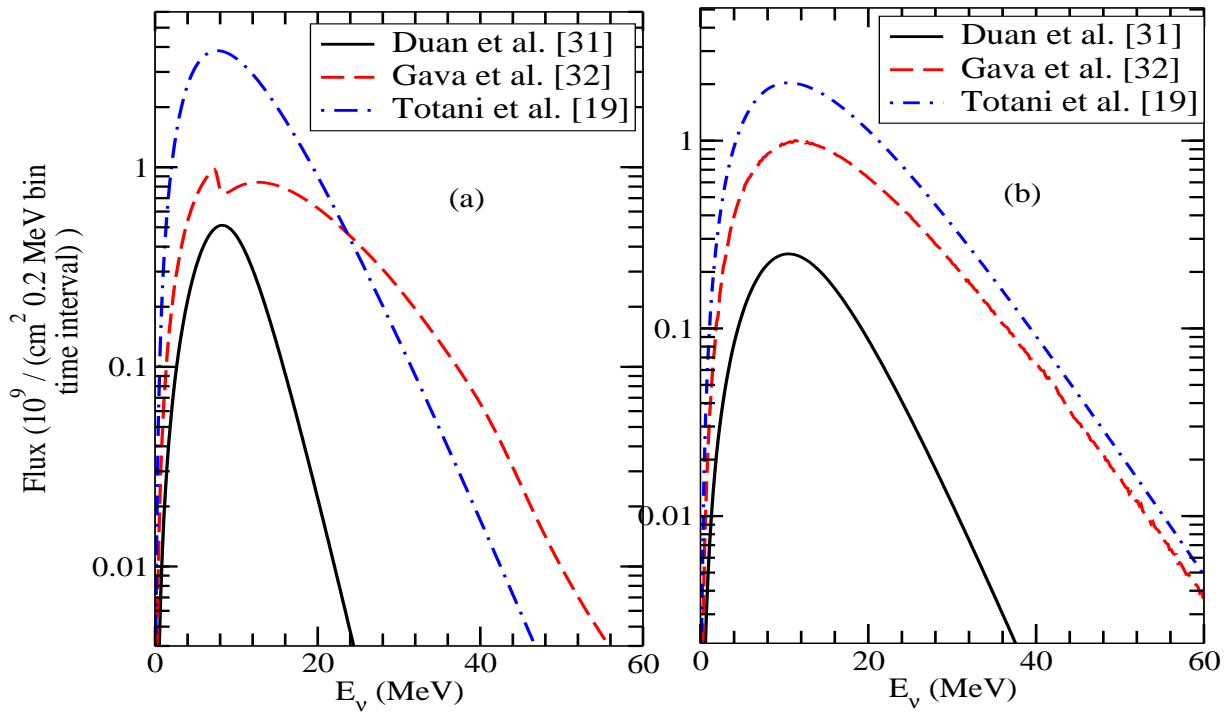}
\caption{Supernova (a) neutrino and (b) antineutrino fluxes simulated by Totani {\it et al.}~\cite{Livermore} (dashed-dotted line), 
Duan {\it et al.}~\cite{FLUX_WEB} (solid line) and Gava {\it et al.}~\cite{Kneller}(dashed line).}
\label{fig:flux_nue_nuebar}
\end{figure}

In the present work, we have taken the numerical flux for $\nu_e$ and $\bar \nu_e$ given by Totani {\it et al.}\cite{Livermore}, 
Duan {\it et al.}\cite{FLUX_WEB} and Gava {\it et al.}\cite{Kneller} as shown in Fig.(\ref{fig:flux_nue_nuebar}). While all of these spectra peak in 
the region of 8 - 12 MeV, their strength and shape are
predicted to be different for the neutrinos/antineutrinos ($(\bar \nu_e)\nu_e$) .
They all have long high energy tail regions which are different from each other and may lead to quite different results for the flux-averaged cross-sections, angular and lepton energy distributions of the 
produced electron (positron) after interaction with the nuclear target in the detector. 
The main aim of the present work is to study 
\begin{itemize}
 \item the nuclear medium effects in the neutrino/antineutrino-nucleus cross-sections in the medium and heavy nuclei proposed to be used in the 
 present and future supernova detectors.
 
 \item the differences in electron (positron) yields, their angular and energy distributions in the present and 
future experiments arising due to the use of different neutrino/antineutrino fluxes when a given nuclear target is used in a detector.
\end{itemize}

Most of the proposed experiments planned to study supernova neutrinos shall be using $^{12}C$, $^{16}O$, $^{40}Ar$, $^{56}Fe$ and $^{208}Pb$ 
as target nuclei in the detectors as shown in Table-\ref{tab:expt}~\cite{Agafonova:2007hn,Borexino,SNO+,Djurcic:2015vqa, An:2015jdp,Bays:2012iy,Hyperk,Icarus,
Anderson:2011ce, Georgia,Katori:2011uq,Strait:2016mof,Smith:2001ns,Hargrove:1996zv,HALO}. In these nuclei, 
it becomes important to study the role of nuclear medium effects on the neutrino/antineutrino-nucleus cross-sections. 
In the energy region of supernova neutrino/antineutrino energies shown in Fig.(\ref{fig:flux_nue_nuebar}), many transitions leading to higher nuclear states in
the final nucleus in addition to the ground state (g.s.) to ground state (g.s.) transition contribute to the cross-section. The cross-sections 
corresponding to exclusive as well as inclusive reactions are generally computed
 using shell model~\cite{Haxton:1987kc, Haxton:1988mw, Engel:1993nn, Engel:2002hg, Hayes:1999ew, Sampaio:2002sh, Suzuki:2006qd}. However, various
 other microscopic approaches like the 
 random phase approximation (RPA)~\cite{Auerbach:1997ay, Singh:1993rg,Kosmas:1996fh,Singh:1998md,Volpe:2000zn, Volpe:2001gy},
 continuum random phase approximation (CRPA)~\cite{Kolbe:1992xu, Kolbe:1995af, Jachowicz:1998fn, Jachowicz:2002rr, Botrugno:2005kn}, combined shell 
 model and random phase approximation CRPA\cite{Auerbach:1997ay,  Kolbe:1999au, Kolbe:2003ys, Vogel:2006sg},
 quasiparticle random phase approximation (QRPA)~\cite{ Lazauskas:2007bs, Cheoun:2010pn, Chasioti:2009fby, Tsakstara:2011zzd, Tsakstara:2012yd},
 projected QRPA~\cite{Samana:2010up}, relativistic quasiparticle RPA~\cite{Paar:2007fi} and 
 relativistic nuclear energy density functional (RNEDF)~\cite{Paar:2012dj} methods have been used to calculate these cross-sections.
  However, in the case of inclusive reactions, Fermi gas models have 
 also been used to calculate these 
 cross-sections~\cite{LlewellynSmith:1971uhs,Smith:1972xh,Wal.75,Gaisser:1986bv,Kuramoto:1989tk}. This model is later extended to the local Fermi gas (LFG) model
 which takes into account the long
 range correlation effects through RPA~\cite{Athar:2005ca, SajjadAthar:2004yf, SajjadAthar:2005ke, Akbar:2015yda, Kosmas:1996fh, Singh:1998md, 
 Nieves:2004wx, Nieves:2017lij,  Singh:1993rg}.

In this work, we present a calculation of the inclusive reactions for supernova neutrino/antineutrino in $^{12}C$, $^{16}O$, $^{40}Ar$, $^{56}Fe$ and 
$^{208}Pb$ induced by the charged current and the neutral current processes. The effect of Coulomb 
distortion of the outgoing lepton produced in the charged current induced reactions is also taken into account using the modified 
effective momentum approximation (MEMA)~\cite{Engel:1997fy}. This work improves upon our earlier work~\cite{Athar:2005ca, SajjadAthar:2004yf, SajjadAthar:2005ke}, 
 by taking into account the modification due to nuclear medium effect consistently up to the order $\frac{\bf q}{M}$ in the 
 evaluation of neutrino-nucleus cross-section.

In section-II, we present in brief the formalism used to describe various nuclear medium effects included in the calculations. In section-III, results 
are presented and discussed. We conclude our findings in section-IV. 

\section{Formalism}
\subsection{Charged Current(CC) Induced Reactions}
The reaction for the charged current neutrino/antineutrino interaction with a nucleus is given by
\begin{eqnarray}\label{quasi_reaction0}
\nu_e + _{Z}^{A}X &\rightarrow& e^{-} +~ _{Z+1}^{A}Y \nonumber\\ 
\bar\nu_e + _{Z}^{A}X &\rightarrow& e^{+} +~ _{Z-1}^{A}Y',
\end{eqnarray}
for which the basic process is
\begin{eqnarray}\label{quasi_reaction}
\nu_e(k) + n(p) &\rightarrow& e^{-}(k^\prime) + p(p^{\prime}) \nonumber\\ 
\bar\nu_e(k) + p(p) &\rightarrow& e^{+}(k^\prime) + n(p^{\prime}),
\end{eqnarray}
where the quantities in the parentheses represent the four momenta of the respective particles.

 The invariant matrix element for the basic processes is given by
\begin{eqnarray}\label{qe_lep_matrix}
{\cal M}=\frac{G_F}{\sqrt{2}}~l_\mu~J^\mu,
\end{eqnarray}
where $G_F$ is the Fermi coupling constant (=1.16639$\times 10^{-5}$ GeV$^{-2}$), 
and the leptonic weak current is given by
\begin{eqnarray}\label{lep_curr}
l_\mu&=&\bar{u}(k^\prime)\gamma_\mu(1 \mp \gamma_5)u(k),
\end{eqnarray}
where -(+) sign is for the neutrino (antineutrino)-induced process.

$J^\mu$ is the hadronic current given by
\begin{equation}\label{had_curr}
J_\mu=cos\theta_c \bar{u}(p')\left[F_1^V(q^2)\gamma_\mu+F_2^V(q^2)i\sigma_{\mu\nu}\frac{q^\nu}{2M} + F_A^V(q^2)\gamma_\mu\gamma_5 + 
F_P^V(q^2)q_\mu\gamma_5\right]u(p),
\end{equation}
where $\theta_C(=13.1^0)$ is the Cabibbo angle, $q^\mu(E_{\nu} - E_{l},~{\bf k}-{\bf k^\prime})$ is the four momentum transfer and $M$ is the nucleon mass.
$F_{1,2}^V(q^2)$ are the isovector vector form factors and $F_A^V(q^2)$, $F_P^V(q^2)$ are the isovector axial vector and pseudoscalar form factors, 
respectively. The isovector vector form factors $F_{1,2}^V(q^2)$ are written in terms of the  
electromagnetic form factors of proton (neutron), i.e., $F_{1}^{p(n)}(q^2)$ and $F_{2}^{p(n)}(q^2)$, respectively. 
 The expressions of $F_{1,2}^V(q^2)$, $F_A^V(q^2)$ and $F_P^V(q^2)$
  used in this work are described in Appendix A-1.
%

The differential cross-section on the free nucleon is given by
\begin{equation}\label{sig0}
\sigma_{0}({\bf q^2, k^\prime, p}) =  \frac{1}{4\pi} \frac{k^2}{E_{\nu} E_{l}} \frac{M^2}{E_{n} E_{p}} \bar\Sigma\Sigma |{\cal M}^{2}| \delta(q_0 + 
E_{n}  - E_{p}),
\end{equation}
where $q_0=E_{\nu_l}-E_l$, and the matrix element square is obtained by using
Eq.(\ref{qe_lep_matrix}) and is given by
\begin{equation}\label{mat_quasi}
{|{\cal M}|^2}=\frac{G_F^2}{2}~{ L}_{\mu\nu} {J}^{\mu\nu}.
\end{equation}
In Eq.(\ref{mat_quasi}), ${ L}_{\mu\nu}$ is the leptonic tensor calculated to be
\begin{eqnarray}\label{lep_tens}
{L}_{\mu\nu}&=&{\bar\Sigma}\Sigma{l_\mu}^\dagger l_\nu=L_{\mu\nu}^{S} \pm i L_{\mu\nu}^{A},~~~~\mbox{with}\\
L_{\mu\nu}^{S}&=&8~\left[k_\mu k_\nu^\prime+k_\mu^\prime k_\nu-g_{\mu\nu}~k\cdot k^\prime\right]~~~~\mbox{and}\nonumber\\
L_{\mu\nu}^{A}&=&8~\epsilon_{\mu\nu\alpha\beta}~k^{\prime \alpha} k^{\beta},
\end{eqnarray}
where the $+$ sign(- sign) is for the neutrino(antineutrino).

The hadronic tensor ${J}^{\mu\nu}$ given by:
\begin{eqnarray}\label{had_tens}
J^{\mu\nu}&=&\bar{\Sigma}\Sigma J^{\mu\dagger} J^\nu,
\end{eqnarray}
where $J^{\mu}$ defined in Eq.(\ref{had_curr}) with Eqs.(\ref{f1v_f2v}), (\ref{fa}) and (\ref{fp}) have been used for the numerical calculations. 
The detailed expression for the hadronic tensor $J^{\mu\nu}$ is given in Ref. \cite{Akbar:2015yda}.
\subsection{Neutral Current(NC) induced reactions}
\subsubsection{Neutral current induced reactions in the presence of strangeness}
The reaction for the neutral current neutrino/antineutrino-nucleus elastic process is given by
\begin{eqnarray}\label{elastic0}
\nu_{l} +\; _{Z}^{A}X \;&\rightarrow& \;\nu_{l} + \;_{Z}^{A}X,\nonumber\\
\bar\nu_{l} + \; _{Z}^{A}X \; &\rightarrow& \;\bar\nu_{l} + \;_{Z}^{A}X,
\end{eqnarray}
for which the basic reaction is given by
\begin{eqnarray}\label{elastic}
\nu_{l}(k) + N(p) &\rightarrow& \nu_{l}(k^\prime) + N(p^\prime);~~N=p~~or~~n,~~\mbox{$l$ = e, $\mu$, or $\tau$,}\nonumber\\
\bar\nu_{l}(k) + N(p) &\rightarrow& \bar\nu_{l}(k^\prime) + N(p^\prime).
\end{eqnarray}
 The matrix element for the reactions shown in Eq.(\ref{elastic}) is given by Eq.(\ref{qe_lep_matrix}). The expression for the leptonic current is given by Eq.(\ref{lep_curr}), 
whereas the hadronic current  ${J^\mu}$ is now given by 
\begin{eqnarray}\label{NC_curr}
J^\mu = \bar{u}(p^\prime)\left[\gamma_{\mu} \tilde F_{1}^{N}(q^2) + 
\frac{i}{2M}\sigma_{\mu \nu}q^{\nu}\tilde F_{2}^{N}(q^2) + \gamma_{\mu}\gamma_{5} \tilde F_{A}^{N}(q^2)\right]u(p),
\end{eqnarray}
where $\tilde F_{1,2}^{N}(q^2)$ and $\tilde F_{A}^{N}(q^2)$ are the vector and axial vector form factors, respectively. The pseudoscalar term 
proportional to $\tilde F_{P}^{N}(q^2)$ does not contribute in this case due to the presence of massless leptons in the initial and final states in the 
 reactions given in Eq.(\ref{elastic}). The expressions for $\tilde F_{1,2}^{N}(q^2)$ and $\tilde F_{A}^{N}(q^2)$ are described in Appendix A-2.

 
 \subsubsection{Neutral current induced reactions in the presence of nonstandard interaction (NSI)}
 Non Standard Interactions in the neutral current sector are predicted in various models proposed to describe the physics beyond the Standard Model (BSM) 
implied by the observation of neutrino oscillations. However, a model-independent phenomenological parameterization of NSI~\cite{Ohlsson:2012kf,
Miranda:2015dra} is generally used to calculate the additional contributions due to these interactions. A deviation from the expected cross-section in the SM is 
the signature of the presence of NSI. The effect of these NSI in supernova physics specially in the detection of supernova signal in terrestrial 
 detectors has been studied by some authors~\cite{Scholberg:2005qs,Papoulias:2013gha,Papoulias:2016edm}. In this description, the matrix element for 
the NC interaction of a neutrino with hadron using an effective phenomenological Lagrangian is given by

\begin{equation}
{\cal{M}} = \frac{G_F}{\sqrt{2}} \sum_{\substack{q=u,d \\ \alpha,\beta=e,\mu,\tau}} [\bar{\nu}_{\alpha} \gamma^{\mu} (1-\gamma_5) \nu_\beta] 
[\epsilon_{\alpha \beta}^{qL} (\bar{q}~ \gamma_{\mu} (1-\gamma_5) q) + \epsilon_{\alpha \beta }^{qR} (\bar{q} ~\gamma_{\mu} (1+\gamma_5) q)]. 
\end{equation}

The $\epsilon_{\alpha \beta}^{qL,qR}$ parameters describe either nonuniversal $(\alpha = \beta)$ or flavor changing $(\alpha\neq\beta)$ interactions. 
The vector and axial vector couplings $ \epsilon_{\alpha \beta}^{V}$ and $ \epsilon_{\alpha \beta}^{A}$ are given in terms of these couplings as 
$\epsilon_{\alpha \beta}^{V} = \epsilon_{\alpha \beta}^{qL} + \epsilon_{\alpha \beta}^{qR}$ and $\epsilon_{\alpha \beta}^{A} = 
\epsilon_{\alpha \beta}^{qL} - \epsilon_{\alpha \beta}^{qR}$. These parameters $\epsilon_{\alpha \beta}^{qL}$ and $\epsilon_{\alpha \beta}^{qR}$ are 
quite poorly constrained by the existing data from the experiments on neutrino scattering induced by $\nu_e$ and $\nu_{\mu}$ on nucleons and nuclei. Making the 
assumption $\epsilon_{\alpha \beta}^{qL} = \epsilon_{\alpha \beta}^{qR} ~~ (\epsilon_{\alpha \beta}^{qA} = 0),$ the present data on high energy 
$\nu_{\mu}$ and $\nu_e$ scattering seem to be consistent with $\epsilon_{ee}^{V} \approx \epsilon_{e \tau}^{V} = 0$ ~\cite{Scholberg:2005qs,
Barranco:2005yy,Davidson:2003ha}. We, therefore, use only $\epsilon_{e \mu}^{V}$ to calculate the NSI contribution to the neutrino-nucleon scattering 
specially for $ ^{208}Pb$ as done by Papoulias and Kosmas ~\cite{Papoulias:2016edm}. It will be interesting to study the NC excitation of $^{208}Pb$ to 
excited states and observe neutron emissions in the proposed HALO detector ~\cite{HALO}. However, in this work, we report on the contribution of NSI to the 
total cross-section in NC induced $\nu_e$ scattering on $^{208}Pb$. For the numerical calculations, we have used 
the parameterization of Papoulias and Kosmas ~\cite{Papoulias:2016edm} for the weak nucleon form factor in the presence of NSI which are described in Appendix A-3. 
\subsection{Cross-Section on nuclear targets}
When the processes given by Eq.~(\ref{quasi_reaction}) or (\ref{elastic}) take place in a nucleus, various nuclear medium effects like 
Pauli blocking, Fermi motion, binding energy corrections and nucleon correlations, etc. come into play. Moreover, in the case of CC reactions, 
the charged lepton produced 
in the final state moves in the Coulomb field of the residual nucleus and affects its energy and momentum. We have taken into account these 
effects which are briefly discussed below:
 \begin{enumerate}
  \item In the standard treatment of Fermi Gas Model applied to neutrino reactions, 
 the quantum states of the nucleons inside the nucleus are filled up to a Fermi momentum $p_{F}$, given by 
$p_{F}= \left[3 \pi^2 \rho\right]^{\frac{1}{3}}$, where $\rho$ is the density of the nucleus. In a nuclear reaction, the momentum of the initial 
nucleon $p$ is therefore constrained to be 
 $p < p_{F}$ and $p^\prime (= |{\bf p} + {\bf q}|) > p_{F}^\prime$, where $p_{F}$ 
 is the Fermi momentum of the initial nucleon target in the Fermi sea, and 
 $p_{F}^\prime$ is the Fermi momentum of the outgoing nucleon.
 The total energies of the initial ($i$) and final ($f$) nucleons are $E_i=\sqrt{{\bf p}^2+M_i^2}$ and $E_f=\sqrt{|{\bf p} + {\bf q}|^2+M_f^2}$.
 In this model, the Fermi momentum and energy are constrained to be determined by the nuclear density which is constant.  
 
 In the LFG model, the Fermi momenta of the initial and final nucleons are not constant, but depend upon the interaction point 
${\bf r}$ and are given by $p_{F_n}(r)$ and $p_{F_p}(r)$ for neutron and proton, respectively, where 
 $p_{F_n}(r)= \left[3 \pi^2 \rho_n (r)\right]^{\frac{1}{3}}$ and $p_{F_p}(r)= \left[3\pi^2 \rho_p (r)\right]^{\frac{1}{3}}$,
$\rho_n (r)$ and $\rho_p (r)$ being the neutron and proton nuclear densities, respectively, and the expressions are given in Appendix-B.

 For neutrino/antineutrino-nucleon scattering from a nuclear target, we define an occupation number $n_N({\bf p, r})$ such that at position 
 ${\bf r}$, where the interaction takes place, the initial nucleon has 
 $n_i({\bf p, r})$=1 for $p < p_{F}(r)$, where $p_{F}(r)$ is the Fermi momentum at position ${\bf r}$. In Appendix-B, we have also 
 shown the Fermi momentum ($p_{F}(r)$) as a function of position (r) for the various nuclei used in this work.
 
  In the local density approximation (LDA), the cross-section($\sigma$) for the $\nu(\bar\nu)$ scattering from a nucleon moving in the 
  nucleus with a momentum ${\bf p}$ is given by
\begin{equation}\label{sigma_lda}
\sigma(q^2, k^\prime) = \int 2 d{\bf r} d{\bf p} \frac{1}{(2\pi)^3} n_i({\bf p}) [1-n_{f}({\bf p} + {\bf q})] \sigma_{0}({\bf q^2, k^\prime, p}),
~~i(f)=n(p) {~~for ~\nu} ~~and~~ p(n) {~~for ~\bar\nu}.
\end{equation}

Instead of using Eq.~\ref{sigma_lda}, we use the methods of many-body field theory~\cite{Fetter} where the reaction cross-section for the process $\nu_e + n 
\to e^- + p$ in a nuclear medium is given 
in terms of the imaginary part of the Lindhard function $U_N(q_0,{\bf q})$ corresponding to the $p-h$ excitation diagram shown in Fig.(\ref{fig:neutrinoselfenergy})~\cite{Singh:1993rg}.
 This imaginary part $U_N(q_0,{\bf q})$ is obtained by cutting the $W(Z)$ self-energy diagram along the horizontal line(Fig.(\ref{fig:neutrinoselfenergy})) and applying the
Cutkowsky rules~\cite{Itzykson}. This is equivalent to replacing the expression
 \begin{equation}
\int \frac{d\bf{p}}{(2\pi)^3}{n_i(\bf{p})} [1-n_{f}({\bf p} + {\bf q})] \frac{M_n M_p}{E_{n}({\bf p}) E_{p}(\bf p+\bf  q)}\delta[q_0+E_n-E_p],
\end{equation}
occurring in Eq.(\ref{sigma_lda}) through Eq.(8) by $-(1/{\pi})$Im${{U_N}(q_0^{\nu(\bar\nu)},{\bf q})}$ which has been discussed in Appendix-C. 

 \item In the charged current reaction, the energy and momentum of the outgoing charged  lepton are modified due 
to the Coulomb interaction of the charged lepton with the final nucleus. The Coulomb distortion effect on the outgoing lepton 
has been taken into account in MEMA~\cite{Engel:1997fy}
in which the lepton momentum and energy are modified by replacing $E_l$ by $E_l+V_c(r)$. The form of Coulomb potential $V_{c}(r)$ considered here 
is~\cite{Preston}:
\begin{equation}
V_{c}(r) = Z_{f}\alpha~ 4\pi\left(\frac{1}{r}\int_{0}^{r}\frac{\rho_{p}(r^\prime)}{Z}r^{\prime 2}dr^{\prime} + \int_{r}^{\infty}\frac{\rho_{p}
(r^\prime)}{Z}r^{\prime}dr^{\prime}\right),
\end{equation}
where $\alpha$ is fine  structure constant(1/137.035), $Z_{f}$ is the charge of the outgoing lepton which is -1 in the case of neutrino and +1 in the case of antineutrino, and 
$\rho_{p}(r)(\rho_n(r))$ is the proton(neutron) density of the final nucleus. 

Incorporation of these considerations results in the modification in the argument of Lindhard function, i.e.,
\[Im U_N (q_{0}^{\nu(\bar\nu)}, {\bf q})~\longrightarrow~Im U_N (q_0^{\nu(\bar\nu)}~-~V_c(r), {\bf q}).\]

 With the inclusion of these nuclear effects, the cross-section $\sigma(E_\nu)$ is written as
 {\footnotesize
\begin{eqnarray}\label{xsection_medeffect}
\sigma(E_\nu)=-2{G_F}^2\cos^2{\theta_c}\int^{r_{max}}_{r_{min}} r^2 dr \int^{{k^\prime}_{max}}_{{k^\prime}_{min}}k^\prime dk^\prime
 \int_{Q^{2}_{min}}^{Q^{2}_{max}}dQ^{2}\frac{1}{E_{\nu_l}^{2} E_l} L_{\mu\nu}J^{\mu\nu}Im U_N (q_0^{\nu(\bar\nu)}-V_c(r), {\bf q}).~~~~
\end{eqnarray}}

\begin{figure}
\begin{center}
\includegraphics[height=.35\textheight,width=0.35\textwidth]{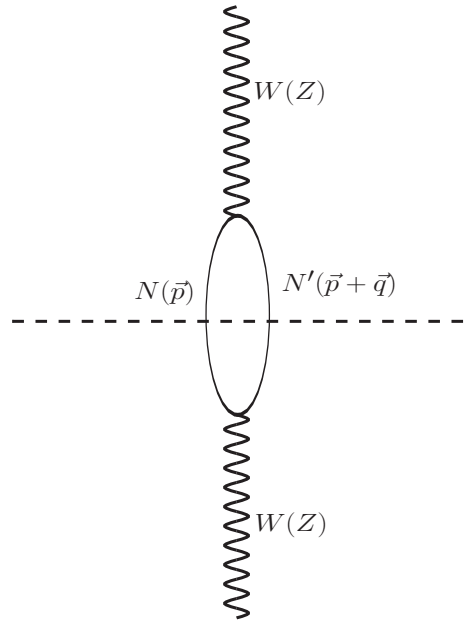}
\caption{Diagrammatic representation of the particle - hole(p-h) excitation induced by $W(Z)$ boson in the large mass limit of intermediate 
 vector boson ($M_{W(Z)}\rightarrow\infty$).}
\label{fig:neutrinoselfenergy}
\end{center}
\end{figure}
\begin{figure}
\begin{center}
\includegraphics[height=.36\textheight,width=0.70\textwidth]{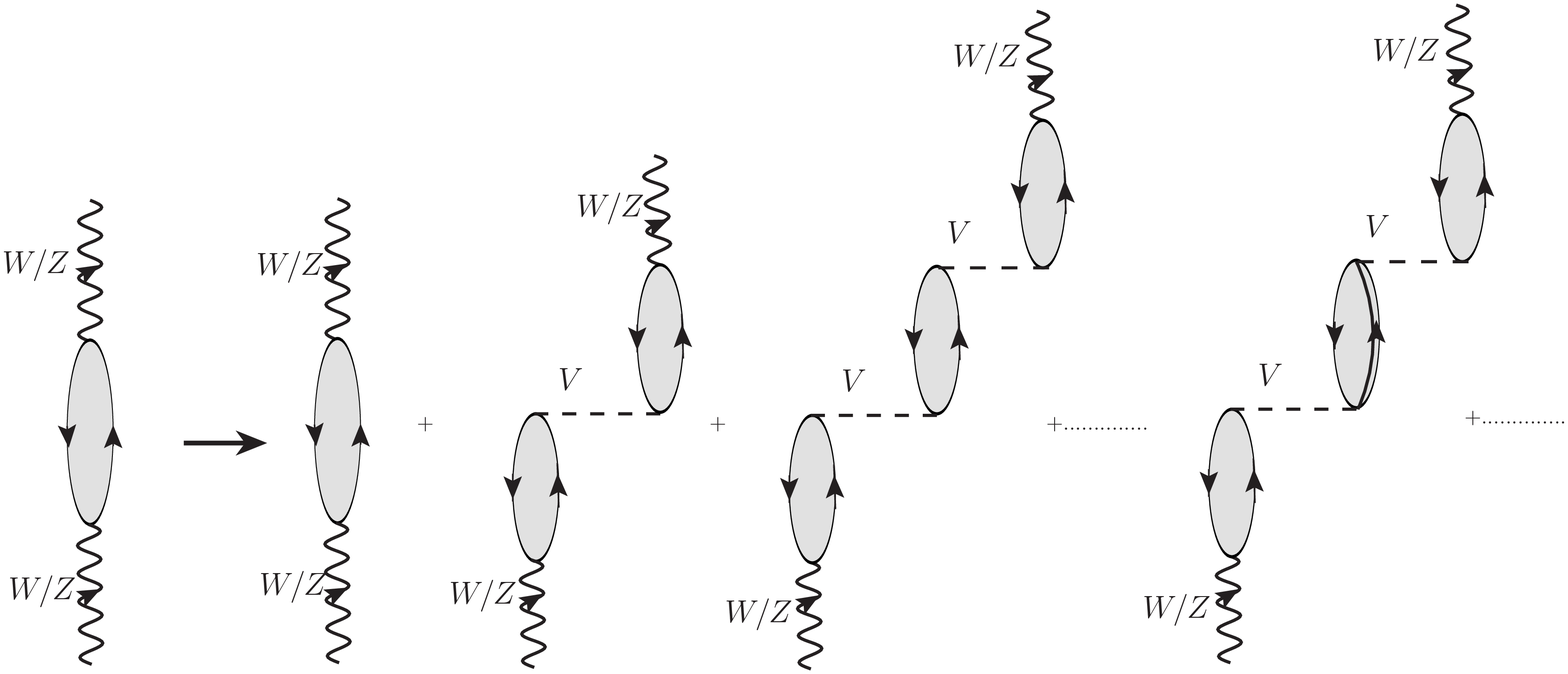}
\caption{RPA effects in the 1p1h contribution to the W/Z self-energy, where particle-hole, $\Delta$-hole, $\Delta$-$\Delta$, etc. excitations 
contribute.}
\label{fg:fig2}
\end{center}
\end{figure}
 \begin{figure}[t]
\includegraphics[height=.65\textheight,width=0.98\textwidth]{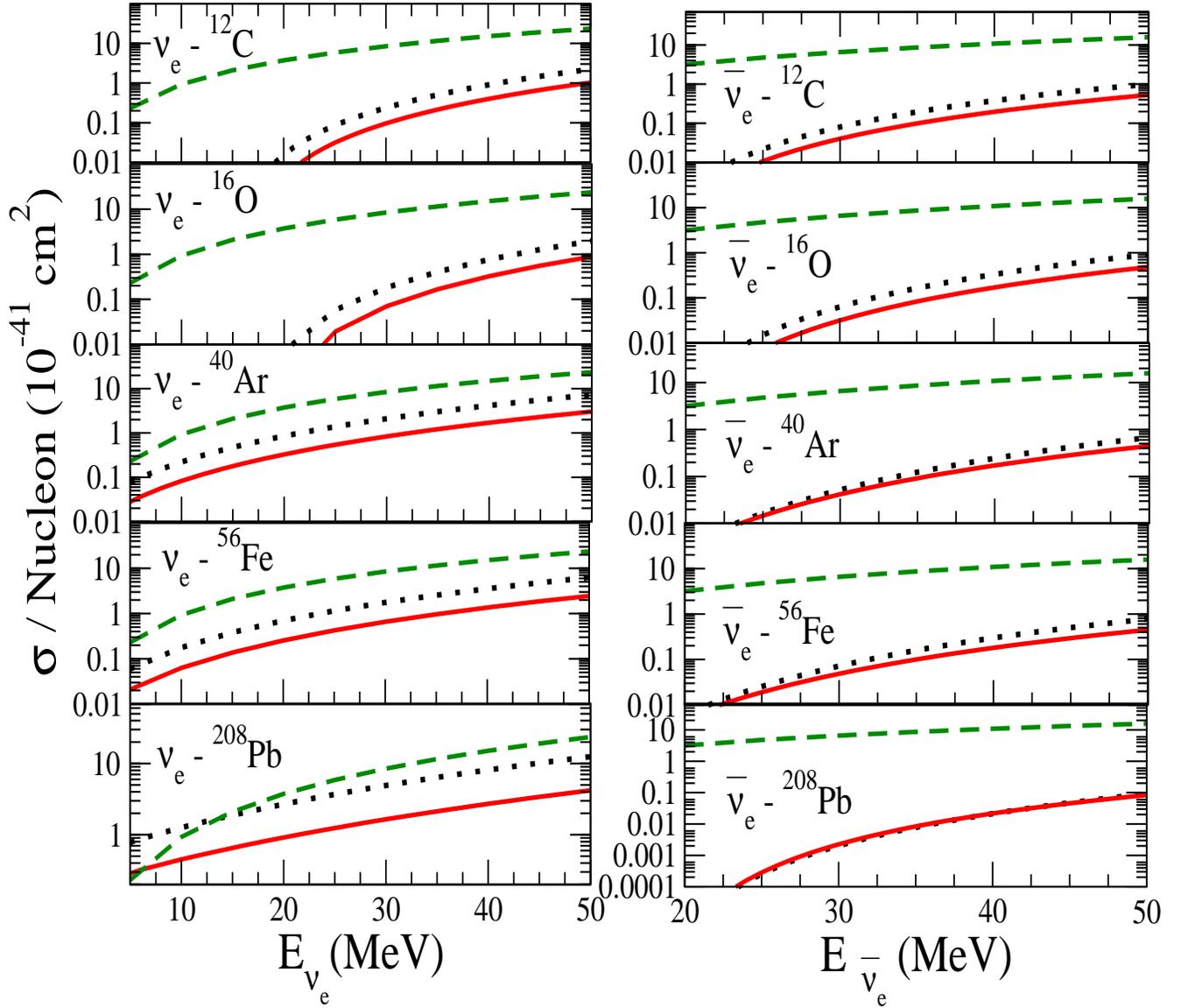}
\caption{(Color online) $\sigma/Nucleon ~ \text{versus} ~ E_{\nu_{e}(\bar \nu_{e})}$ for the CCQE scattering induced by $\nu_e$ (left panel) and $\bar \nu_e$
(right panel) in $^{12}C$, $^{16}O$, $^{40}Ar$, $^{56}Fe$ and $^{208}Pb$. The results are presented for the cross-section obtained using 
Eq.(\ref{xsection_medeffect}) (without RPA) and Eq.(\ref{xsection_final}) (with RPA) shown here by dotted line and solid line, respectively.  
The dashed line is the result for $\nu_{e}(\bar \nu_e)$ scattering off free nucleon target.}
\label{fig:cc_100_log}
\end{figure}
\item  In the nucleus, the strength of the electroweak couplings may change from their free nucleon values due to the presence of 
strongly interacting nucleons. Conservation of vector current (CVC) forbids any change in the charge coupling while the magnetic 
and axial vector couplings are likely to change from their free nucleon values. There exists considerable work in understanding the 
 quenching of magnetic moment and axial charge in nuclei due to the nucleon-nucleon correlations.
 In our approach, these are reflected in the modification  of nuclear response in the longitudinal and transverse channels leading to some reduction. 
 We calculate this reduction in the vector-axial(VA) and axial-axial(AA) response functions due to the long range nucleon-nucleon correlations treated in the RPA, diagrammatically shown in Fig.(\ref{fg:fig2}). 
\begin{figure}
\includegraphics[height=.7\textheight,width=0.75\textwidth]{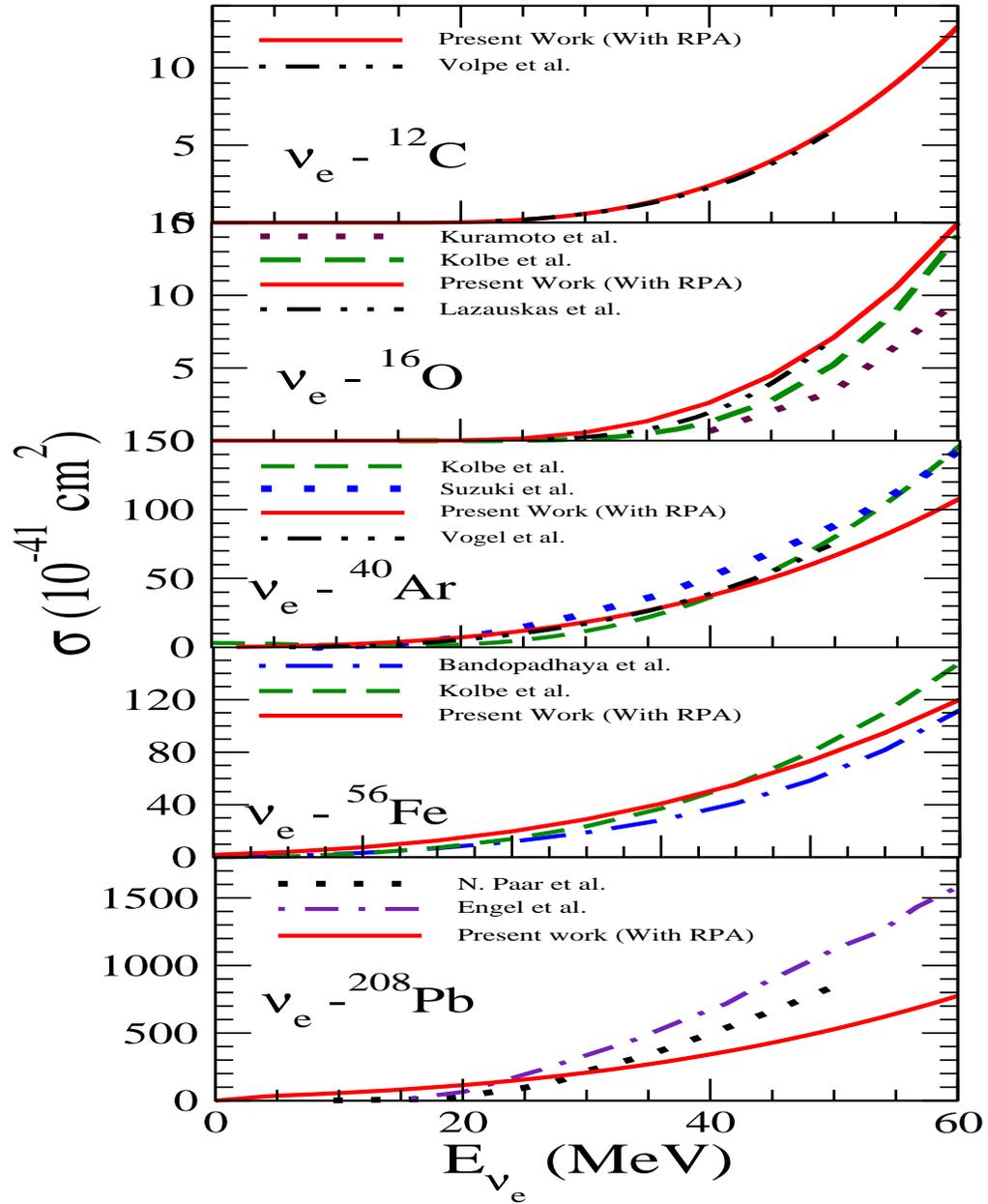}  
\caption{Comparison for $\sigma$ \text{versus} $E_{\nu_e}$ in various nuclei. The results are presented for the 
cross-section obtained with (solid line) RPA correlations. For $\nu_{e}-^{12}C$ scattering results have been compared with the results of
Volpe {\it et al.}~\cite{Volpe:2000zn} (dashed-double dotted line), $\nu_{e}-^{16}O$ scattering results are compared with Kuramoto {\it et al.}
\cite{Kuramoto:1989tk} (dotted line), Kolbe {\it et al.}\cite{Kolbe:2002gk} (dashed line) and Lazauskas {\it et al.}~\cite{Lazauskas:2007bs}(dashed-double dotted line), 
 $\nu_{e}-^{40}Ar$ scattering results are compared with 
Kolbe {\it et al.}\cite{Kolbe:2003ys}(dashed line), Vogel {\it et al.}\cite{Volpe:2007qx} (dashed-double dotted line) and Suzuki {\it et al.}~\cite{Suzuki:2012ds} (dotted line), $\nu_{e}-^{56}Fe$ scattering results are compared with Kolbe {\it et al.}~\cite{Kolbe:2000np}
(dashed line) and Bandopadhayay {\it et al.}~\cite{Bandyopadhyay:2016gkv} (dashed-dotted line) and results for $\nu_{e}-^{208}Pb$ are compared with the 
results of Engel {\it et al.}\cite{Engel:2002hg} (dashed-dotted line) and Paar {\it et al.}~\cite{Paar:2008zza} (dotted line).}
\label{fig:sig_compare}
\end{figure}
The weak nucleon current described by Eq.(\ref{had_curr}) gives in the nonrelativistic limit, terms like $F_A {\bm \sigma}\tau_+$ 
and $i F_2^V \frac{{\bm \sigma}\times {\bf q}}{2M}\tau_+$ which generate spin-isospin transitions in nuclei. While the term 
$i F_2^V \frac{{\bm \sigma}\times {\bf q}}{2M}\tau_+$ couples with the transverse excitations, the term  $F_A {\bm \sigma}\tau_+$ 
couples with the transverse as well as the longitudinal channels. These channels produce different RPA responses in the longitudinal and 
transverse channels due to the different NN potential in these channels when the diagrams of Fig.(\ref{fg:fig2}) are summed up. 
As a consequence, a term proportional to $F^2_A \delta_{ij}$ in $J^{ij}$ is replaced by $J^{ij}_{RPA}$ as \cite{Akbar:2015yda}:
\begin{equation}\label{f2a_rpa}
J^{ij}\rightarrow J^{ij}_{RPA}= F^2_A{Im U_N}\left[\frac{{\bf{\hat{q_i}}{\hat{q_j}}}}{1-U_NV_l}+\frac{\delta_{ij}-{\bf{\hat{q_i}}{\hat{q_j}}}}
{1-U_NV_t}\right],
\end{equation}
\begin{figure}[t]
\includegraphics[height=.3\textheight,width=0.95\textwidth]{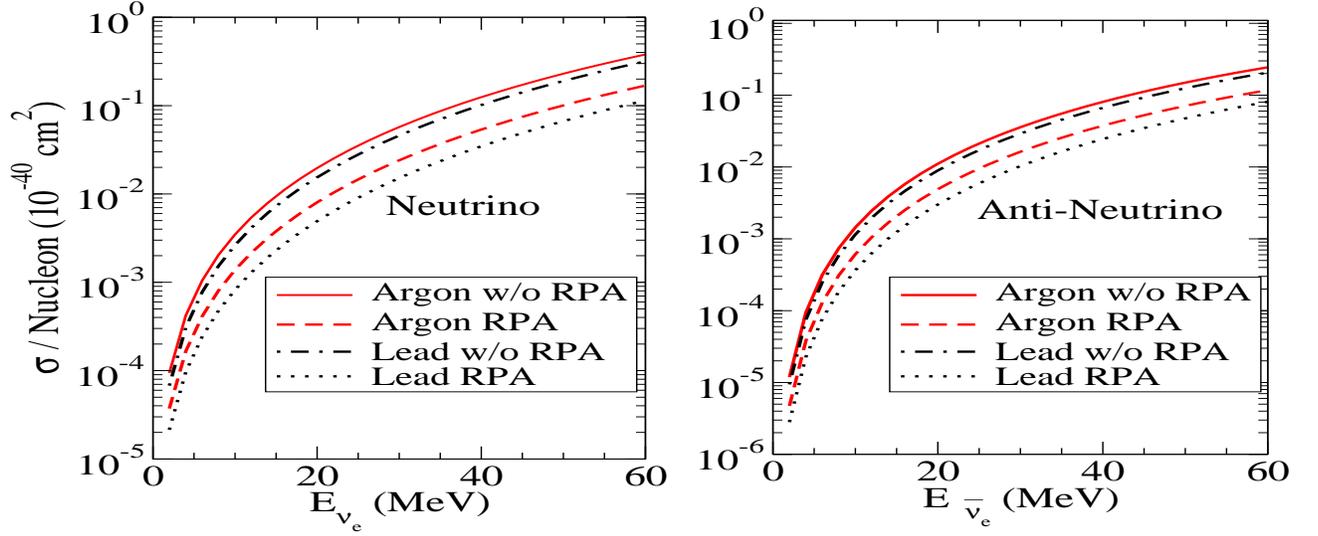}
\caption{$\sigma/Nucleon~ \text{versus} ~E_{\nu(\bar \nu)}$ for neutral current neutrino- (left panel) and antineutrino (right panel)-induced processes 
in argon and lead.}
\label{fig:nc_xsection}
\end{figure} 
where the first and second terms show the modification in $J^{ij}$ in longitudinal and transverse channels. 
In Eq.(\ref{f2a_rpa}), $V_l$ and $V_t$ are the longitudinal and transverse parts of the nucleon-nucleon potential calculated using $\pi$ and $\rho$ 
exchanges and are given by 
\begin{eqnarray}\label{longi_part}
V_l(q) = \frac{f^2}{m_\pi^2}\left[\frac{q^2}{-q^2+m_\pi^2}{\left(\frac{\Lambda_\pi^2-m_\pi^2}{\Lambda_\pi^2-q^2}\right)^2}+g^\prime\right],\nonumber\\
V_t(q) = \frac{f^2}{m_\pi^2}\left[\frac{q^2}{-q^2+m^2_\rho}{C_\rho}{\left(\frac{{\Lambda_\rho}^2-m^2_\rho}{{\Lambda_\rho}^2-q^2}\right)^2}+g^\prime
\right].\end{eqnarray}
$\frac{f^{2}}{4\pi}$ = 0.8, $\Lambda_\pi$ = 1.3 GeV, $C_\rho$ = 2, $\Lambda_\rho$ = 2.5 GeV, $m_\pi$ and $m_\rho$ are the pion and rho meson masses, 
and $g^\prime$ is the Landau-Migdal parameter taken to be $0.7$ 
which has been used quite successfully to explain many electromagnetic and weak processes in nuclei \cite{Singh:1998md, Gil:1997bm, Carrasco:1989vq}.

The effect of the $\Delta$ degrees of freedom in the nuclear medium is included in the calculation of the RPA response by considering 
the effect of ph-$\Delta$h and $\Delta$h-$\Delta$h excitations. 
This is done by replacing $U_N \rightarrow U_N^{\prime}=U_N+U_\Delta$, where $U_\Delta$ is the Lindhard function for the
$\Delta$h excitation in the nuclear medium. The expressions for $U_N$ and $U_\Delta$ are taken from Ref.\cite{Oset1}. 
The different couplings of $N$ and $\Delta$ are incorporated in $U_N$ and $U_\Delta$ and then the same interaction strengths 
($V_l$ and $V_t$) are used to calculate the RPA response. These effects have been discussed by Nieves {\it et al.} \cite{Nieves:2004wx} 
 as well as by Athar {\it et al.}\cite{SajjadAthar:2005ke}.
\begin{figure}
\includegraphics[height=.3\textheight,width=0.95\textwidth]{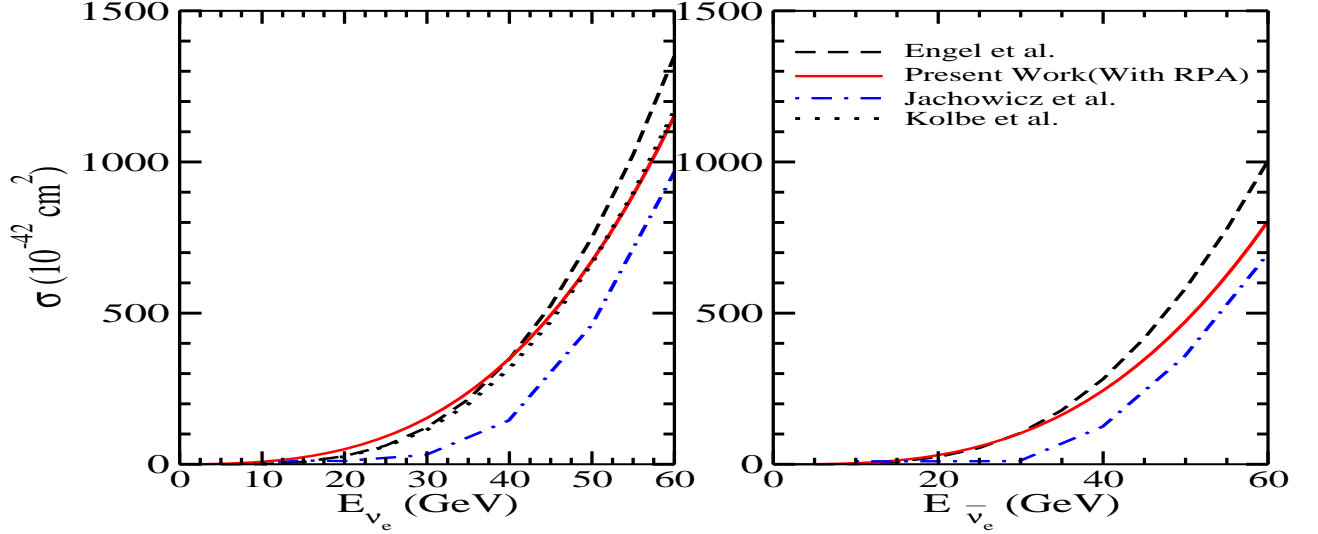} 
\caption{Comparison for $\sigma_{\nu}$ for neutral current neutrino- (left panel) and antineutrino- (right panel)
induced processes in $^{208}Pb$. The results presented here are obtained in the local Fermi gas model with (solid line) RPA correlations.
 The results of Engel {\it et al.}\cite{Engel:2002hg} (dashed line), Kolbe {\it et al.}~\cite{Kolbe:2000np} (dotted line) and 
 Jachowicz {\it et al.}~\cite{Jachowicz:2002hz} (dashed-dotted line) are also shown.}
\label{fig:NC_compare}
\end{figure}
\begin{figure}
\includegraphics[height=.3\textheight,width=0.95\textwidth]{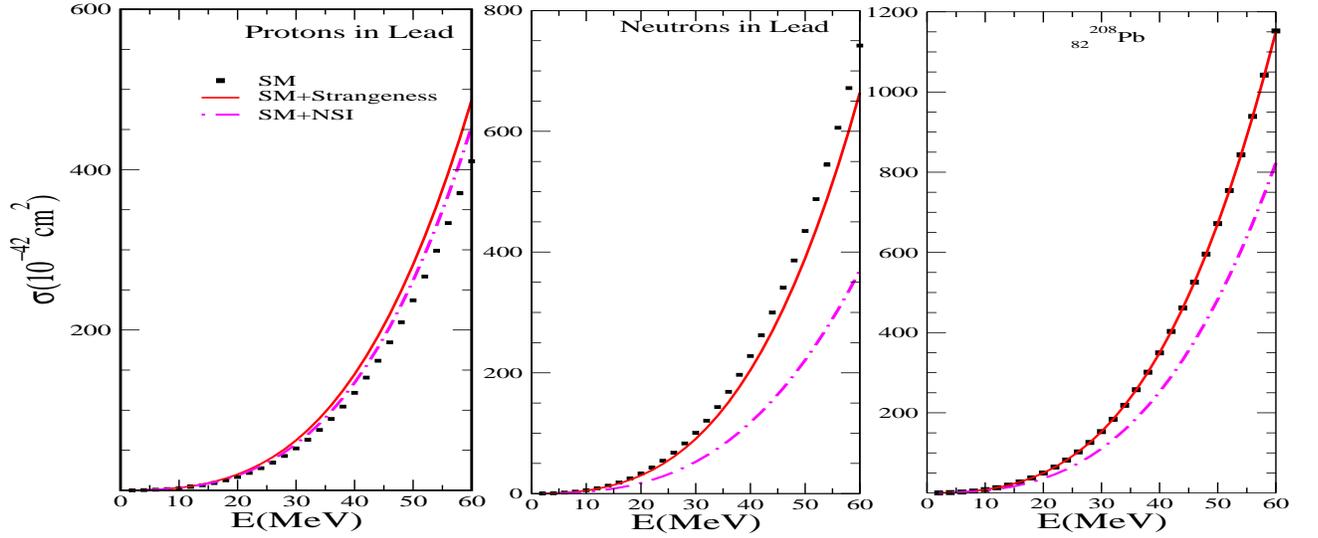} 
\caption{Total scattering cross-section $\sigma_{\nu}$ obtained using LFGM with RPA effect for neutral current neutrino-induced 
processes for (i) the protons in $^{208}Pb$ (left panel), (ii) the neutrons in $^{208}Pb$ (central panel), and (iii) the nucleons in $^{208}Pb$ nuclear target (right panel).
 The results are presented in the Standard Model (dotted line) using Eq.(\ref{NC_curr}) with the form factors defined in Eq.(\ref{ncff1}) with $F_{1,2}^{s}(q^2)$=0 and
 Eq.(\ref{fanc}) with $\Delta$s=0, results in the Standard Model by taking $\Delta$s=-0.12 in Eq.(\ref{str-par2}) (solid line), and 
 results in the Standard Model with the nonstandard interaction using Eq.~(\ref{nsi-expr}) with $F_{1,2}^{s}(q^2)$=0 and $\Delta$s=0 (dashed-dotted line).}
\label{fig:NC_compare1}
\end{figure}
With the incorporation of these nuclear medium effects, the expression for the total scattering cross-section $\sigma(E_\nu)$ is given 
by Eq.(\ref{xsection_medeffect}) with $J^{\mu \nu}$ replaced by $J^{\mu \nu}_{RPA}$(defined in Eq. (\ref{f2a_rpa})) i.e.,
\begin{eqnarray}\label{xsection_final}
\sigma(E_\nu)&=&-2{G_F}^2 a^{2} \int^{r_{max}}_{r_{min}} r^2 dr \int^{{k^\prime}_{max}}_{{k^\prime}_{min}}k^\prime dk^\prime
 \int_{Q^{2}_{min}}^{Q^{2}_{max}}dQ^{2}\frac{1}{E_{\nu_l}^{2} E_l}  L_{\mu\nu}J^{\mu \nu}_{RPA} Im{U_N}(q_0^{\nu(\bar\nu)}-V_c(r)),~~~~~~
\end{eqnarray}
where $J^{\mu \nu}_{RPA}$ is the hadronic tensor with its various components modified due to long range correlation effects treated in RPA as it is 
shown in Eq.(\ref{f2a_rpa}) for the leading term proportional to $F_A^2$.
In Eq.(\ref{xsection_final}), $a= cos \theta_{c}$ for charged current reaction. For the neutral current reactions $a= 1$ with the Lindhard function 
calculated without the Coulomb potential $V_c(r)$.  The full expression for $J^{\mu \nu}_{RPA}$ is given in the Appendix of Ref.~\cite{Akbar:2015yda}.
\end{enumerate}

\begin{figure}
\includegraphics[height=.75\textheight,width=0.99\textwidth]{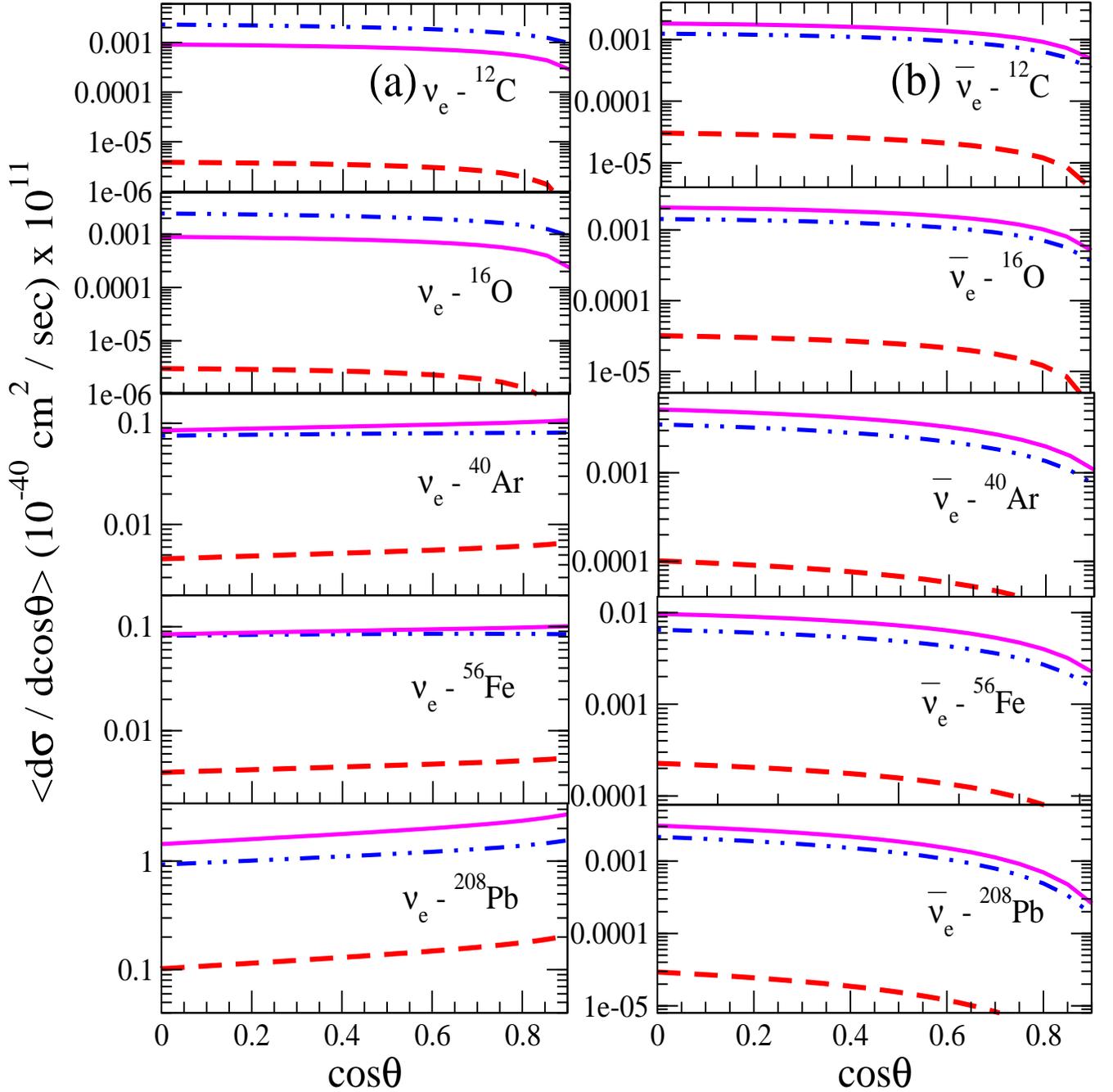}
\caption{Angular distribution of lepton, averaged over supernova (a) neutrino (left panel) and (b) antineutrino (right panel) fluxes.
 All the results are presented in the units of $10^{-40}cm^{2}/sec$ and is to be multiplied by $10^{11}$(corresponding to Fig.(1)). 
 These results are obtained 
 with RPA effects. Here solid line, dashed line, dashed double-dotted line show the angular distributions averaged over the fluxes 
 simulated by Totani {\it et al.} \cite{Livermore}, Duan {\it et al.} \cite{FLUX_WEB}, Gava {\it et al.} \cite{Kneller}.}
\label{fig:cc_dsdcost}
\end{figure}

\begin{figure}
\includegraphics[height=.7\textheight,width=0.99\textwidth]{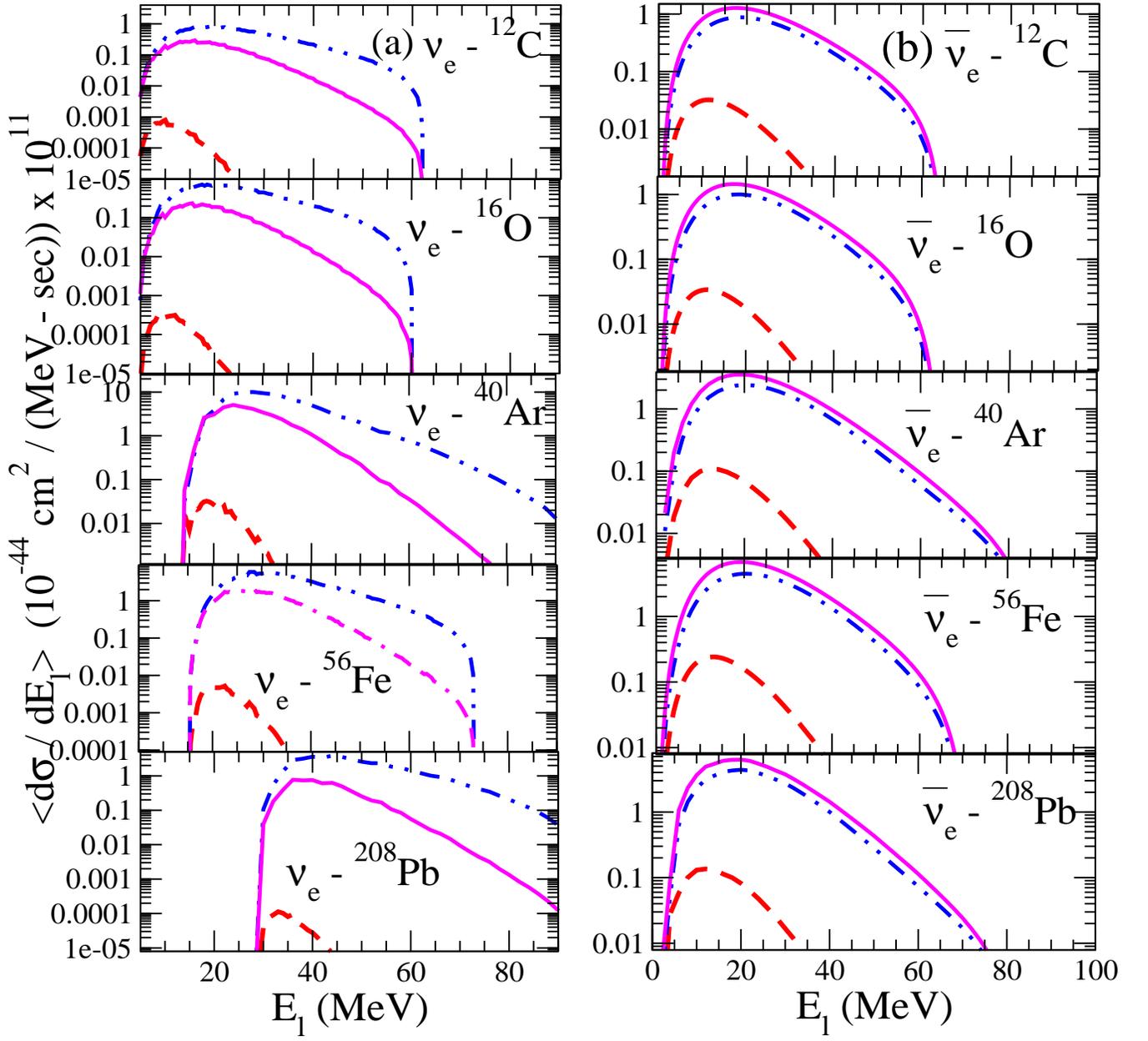}
\caption{Lepton energy distribution averaged over supernova (a) neutrino (left panel) and (b) antineutrino (right panel) fluxes. 
All the results are presented in the units of $10^{-44}cm^{2}/(MeV-sec)$ and is to be multiplied by $10^{11}$(corresponding to Fig.(1)).
 Lines and points have the same meaning as in Fig.~(\ref{fig:cc_dsdcost}).}
\label{fig:cc_dsdel}
\end{figure}

\subsection{Flux-averaged cross-section}
\subsubsection{Low energy accelerator neutrinos}

Using the expressions for the cross-section $\sigma(E_\nu)$ in Eqs.(\ref{xsection_medeffect}) and (\ref{xsection_final}), the flux-averaged cross-sections $\langle\sigma\rangle$ are obtained as
\begin{equation}\label{fluxavg_LSND_michel}
\langle\sigma\rangle = \frac{\int_{E_{\nu}^{min}}^{E_{\nu}^{max}} \sigma(E_{\nu}) \mathit{f}(E_{\nu}) dE_{\nu}}{\int_{E_{\nu}^{min}}^{E_{\nu}^{max}}  
\mathit{f}(E_{\nu}) dE_{\nu}},
\end{equation}
where $\mathit{f}(E_{\nu})$ is the spectrum of the neutrino from a low energy source like Michel spectrum from muon decay at rest ($\mu$DAR) or 
pion decay in flight ($\pi$DIF) at LAMPF.

$\nu_e$ Michel spectrum (for $\mu$DAR) is given as
\begin{equation}\label{michel}
\mathit{f}(E_{\nu_e}) = \dfrac{12}{E_{0}^{4}} E_{\nu_e}^{2} (E_{0} - E_{\nu_e}) \nonumber
\end{equation}
with $E_{0}$ = 52.8 MeV. For $\nu_\mu$ flux from the $\pi$DIF corresponding to LSND experiment~\cite{Albert:1994xs},
 we have used the flux given in  Ref.~\cite{Nieves:2004wx} with $E_{\nu}^{min}$ = 123.1 MeV and $E_{\nu}^{max}$ = 300 MeV.
 
 \begin{table}
\begin{tabular}{|c|c|c|c|}\hline\hline
Process & Experimental & Present work& Other Calculations\\ \hline
$\nu_{e}-^{12}C$ ($10^{-42} cm^{2}$)&  $15 \pm 1 \pm 1 $(KARMEN)\cite{Bodmann:1994py} & $14.9 $ &  15.6 \cite{Kolbe:1995af}  \\
&$14.8 \pm 0.7 \pm 1.1 $ (LSND)\cite{Athanassopoulos:1997rm}  &  & 12.9 - 22.7 \cite{Auerbach:1997ay}\\    
&  $14.1 \pm 2.3 $(LAMPF)\cite{Krakauer:1991rf}   &  & 14 \cite{Nieves:2004wx} \\
&    &  & $13.8 \pm 0.4$ \cite{Nieves:2017lij} \\
&  $15 \pm 1 $(LSND)\cite{Athanassopoulos:1997rm}   &  &$12.14$ \cite{Paar:2007fi}  \\
 $\nu_{e}-^{56}Fe$($10^{-42} cm^{2}$) & $256 \pm 108 \pm 43$\cite{SajjadAthar:2005ke}  & 300 & 240 \cite{Kolbe:2000np}\\
  &  &  & 264.6 \cite{Samana:2008pt}\\
  &  &  & 197.3 \cite{Samana:2008pt}\\
 &  &  & 352.0 \cite{Lazauskas:2007bs}\\
 &  &  & 140.0 \cite{Paar:2007fi}\\
$\nu_{\mu}-^{12}C$ ($10^{-40} cm^{2}$) & $8.3 \pm 0.7 \pm 1.6$ \cite{Albert:1994xs}  & 14.4 & 19.3 - 20.3 \cite{Kolbe:1995af}\\ 
 &  $11.2 \pm 0.3 \pm 1.8$ \cite{Athanassopoulos:1997rn}  &  & 13.5 - 15.2 \cite{Auerbach:1997ay}\\
 & $10.6 \pm 0.3 \pm 1.8$ \cite{Auerbach:2002iy}  &  & 11.9  \cite{Nieves:2004wx}\\
 &&&$19.59$ \cite{Paar:2007fi}\\
  &  &  & $9.7 \pm 0.3$ \cite{Nieves:2017lij}\\\hline
\end{tabular}
\caption{Flux averaged cross-section for $\nu_e$  and $\nu_\mu$ induced processes on $^{12}C$  and $^{56}Fe$ nuclear targets.}\label{flux_averaged_xsection}
\end{table}
 
\subsubsection{Supernova neutrinos/antineutrinos}
We define the flux-averaged cross-section for the supernova neutrinos/antineutrinos as
\begin{equation}\label{fluxavg}
\langle\sigma\rangle_{\nu(\bar\nu)} = \int \sigma(E_{\nu(\bar\nu)}) \mathit{f}(E_{\nu(\bar\nu)}) dE_{\nu(\bar\nu)}, 
\end{equation}
 where $\mathit{f}(E_{\nu})$ is the flux for the supernova neutrinos/antineutrinos. 
We have parameterized the neutrino/antineutrino fluxes given in the numerical tables of 
 Totani {\it et al.} \cite{Livermore}, Duan {\it et al.} \cite{FLUX_WEB} and Gava {\it et al.} \cite{Kneller} using B-spline function~\cite{bspline} 
 and used them in the present calculations.
  The event rates for the charged lepton production has been calculated using the expression
\begin{equation}\label{events}
\text{Event rate}  = \langle\sigma\rangle \times \Delta t \times N_{\text{target}},
\end{equation}
where $\langle\sigma\rangle$ is the flux-averaged cross-section defined in Eq.(\ref{fluxavg}), $\Delta t$ is the time interval (which we have taken 
as 1 s) and $N_{\text{target}}=MN_A$ 
is the number of nucleons with $M$ as the mass of the target material which we have taken equal to 1 kT in our numerical calculations
 and $N_A$ is the Avogadro's number = 6.023$\times$10$^{26}$ kmol$^{-1}$.\\

\section{Results and Discussion}
The numerical calculations have been performed for the energy dependence of $\nu(\bar\nu)-A$ total scattering cross-section $\sigma(E)$
 as well as the angular and energy distribution of the electrons (positrons) for the supernova neutrino
  spectra given in Fig.(1). These results have been presented using the expression of the cross-section given in  Eq.(\ref{xsection_final}) with RPA 
  effect.
 The averaged cross-section is defined in Eq.(\ref{fluxavg_LSND_michel}) for the low energy 
accelerator neutrinos and in Eq.(\ref{fluxavg}) for the supernova neutrinos. The nuclear densities given in Eqs.(\ref{MHO}) \& (\ref{2pF}), with the 
parameters shown in Table-IV are used in these calculations.

In subsection~\ref{suba}, we first show our results for the energy dependence of the total 
cross-section in the low energy region of neutrinos for $^{12}C$, $^{16}O$, $^{40}Ar$, $^{56}Fe$ and $^{208}Pb$ and discuss the nuclear medium effects,
and compare them with other calculations.
 In subsection~\ref{subb}, we have presented the event rates for the charged lepton production obtained using various supernova $\nu(\bar\nu)$ fluxes
given by Totani {\it et al.}~\cite{Livermore}, Duan {\it et al.}~\cite{FLUX_WEB} and Gava {\it et al.}~\cite{Kneller}. 
 In subsections~\ref{subc} and \ref{subd}, we present, respectively, the results for angular distribution and energy distribution of electrons (positrons) produced in the charged 
 current reactions using the above fluxes.  
 \begin{table}\label{event_rates_nue_nuebar}
\begin{tabular}{|c|c|c|c|c|c|c|}\hline\hline
& \multicolumn{2}{|c|}{Duan {\it et al.} \cite{FLUX_WEB}} & \multicolumn{2}{|c|}{Gava {\it et al.} \cite{Kneller}} & \multicolumn{2}{|c|}{Totani 
{\it et al.} \cite{Livermore}} \\\hline
& $\nu_e$ & $\bar\nu_e$&  $\nu_e$ & $\bar\nu_e$&  $\nu_e$ & $\bar\nu_e$\\ \hline
$^{12}C$ & 0.04 & 0.54 & 25 & 13 & 9.52 & 19 \\
$^{16}O$ & 0.03 & 0.52 & 26 & 15 & 9.21 & 22\\
$^{40}Ar$ & 60 & 2 & 907 & 38 & 1057 & 54 \\
$^{56}Fe$ & 60 & 2.4 & 995 & 70 & 1102 & 103\\
$^{208}Pb$ & 1508 & 0.44 & 12103 & 23  & 19268 & 31\\ \hline \hline
\end{tabular}
\caption{Comparison of the event rates obtained for $\nu_{e}(\bar\nu_e)$ induced scattering from $^{12}C$, $^{16}O$, $^{40}Ar$, $^{56}Fe$ and 
$^{208}Pb$ nuclear
targets using 1 kT of target material. These events are calculated in the local Fermi gas model with RPA effect using supernova flux given by
Totani {\it et al.}~\cite{Livermore}, Duan {\it et al.}~\cite{FLUX_WEB} and Gava {\it et al.}~\cite{Kneller}.}
\end{table}
\subsection{Total cross-section}\label{suba}

We present and discuss the effect of nuclear medium including nucleon correlations on the cross-sections  using LFG model for the charged current as 
well as for the neutral current induced reactions.  
\begin{enumerate}[label=\Roman*.]

\item In Fig.(\ref{fig:cc_100_log}), we have shown the results for the total scattering cross-section per nucleon in $^{12}C$, $^{16}O$, $^{40}Ar$, 
$^{56}Fe$ and $^{208}Pb$ for 
$\nu_{e}(\bar\nu_{e})$-induced charged current quasielastic process. We observe that: 
\begin{enumerate}[label=\roman*.]
\item For the case of $\nu_e$-induced scattering on low mass targets like $^{12}C$ and $^{16}O$ (left panel), 
the reduction in the cross-section due to the nuclear medium effects like Pauli blocking and Fermi motion is around $98\%$ at $E_{\nu}$= 20 MeV and 
$90\%$ at $E_{\nu}$= 50 MeV in comparison to the free $\nu_{e}-N$ cross-section. The inclusion of RPA correlation further reduces the cross-section 
by $66\%$ at $E_{\nu}$= 20 MeV and by $55\%$ at $E_{\nu}$= 50 MeV. 

In the case of charged current $\nu_e$-induced processes in high mass target like $^{40}Ar$($^{56}Fe$) nucleus, the reduction in the cross-section
due to the nuclear effects like Pauli blocking and Fermi motion is around $68(80)\%$ at $E_{\nu}$= 20 MeV and $64(72)\%$ at $E_{\nu}$= 50 MeV in comparison 
to the results obtained for the free nucleon case. The inclusion of RPA correlations results in a further reduction in the cross-section, which is $60(64)\%$ at 20 MeV and 
$58(60)\%$ at 50 MeV while in 
$^{208}Pb$, this reduction is around $14\%$ at $E_{\nu}$= 20 MeV and $40\%$ at $E_{\nu}$= 50 MeV in comparison to the results obtained for the free nucleon which becomes 
$66\%$ at 20 MeV and $65\%$ at 50 MeV when RPA correlation is also incorporated.

\item For the case of $\bar\nu_e$-induced scattering on $^{12}C$ and $^{16}O$ (right panel), the cross-section almost reduced by half at $E_{\nu}$= 20 
MeV and by $95\%$ at $E_{\nu}$= 50 MeV in comparison to the results obtained for the free nucleon case. The additional reduction in the cross-section due to the RPA correlation is $50 
\%$ at 20 MeV, which becomes $45 \%$ at 50 MeV. 

In the case of charged current $\bar\nu_e$-induced process in  $^{40}Ar$($^{56}Fe)$, the reduction in the cross-section is almost by half at
$E_{\nu}$= 20 MeV and by $95\%$ at $E_{\nu}$= 50 MeV in comparison to the free $\nu_{e}-N$ cross-section. 
The cross-section further reduces by 16(14)$\%$ at 20 MeV, and 30(40)$\%$ at 50 MeV due to RPA correlations.  
In the case of $^{208}Pb$ nuclear target, the cross-section reduces by half at $E_{\nu}$= 20 MeV and $\sim 98\%$ at $E_{\nu}$= 50 MeV in comparison to 
the free $\nu_{e}-N$ cross-section. 
\end{enumerate}

\item  In Fig.(\ref{fig:sig_compare}), we have compared our results of 
 $\sigma$ for $\nu_e$ charged current induced scattering  with the results of $\sigma$ calculated by Volpe {\it et al.}~\cite{Volpe:2000zn} in $^{12}C$, 
  Kuramoto {\it et al.}\cite{Kuramoto:1989tk},  Kolbe {\it et al.}\cite{Kolbe:2002gk} and Lazauskas {\it et al.}~\cite{Lazauskas:2007bs} 
  in $^{16}O$, Kolbe {\it et al.}\cite{Kolbe:2003ys}, Vogel {\it et al.}\cite{Volpe:2007qx} and Suzuki {\it et al.}~\cite{Suzuki:2012ds} in $^{40}Ar$, 
  Bandopadhayay {\it et al.}\cite{Bandyopadhyay:2016gkv} and Kolbe {\it et al.}\cite{Kolbe:2000np} in $^{56}Fe$  and Engel {\it et al.}
  \cite{Engel:2002hg} and 
  Paar {\it et al.}\cite{Paar:2008zza} in $^{208}Pb$. 
 
\item In Fig.(\ref{fig:nc_xsection}), we have presented the results for $\sigma$ in $^{40}Ar$ and $^{208}Pb$ nuclear targets, as a function of $E_{\nu(\bar \nu)}$,
 for the neutral current induced 
processes using the expression of $J^{\mu}$ given in Eq.(\ref{NC_curr}) with the form factors defined in Eq.(\ref{ncff1}) with $F_{1,2}^{s}(q^2)$=0 .
We find that the inclusion of RPA correlations reduces the cross-section by around 60$\%$ at 10 MeV and 55$\%$ at 40 MeV in $^{40}Ar$, while in 
$^{208}Pb$, this reduction is a bit larger, for example, at 
10 MeV, this reduction is 70$\%$ which becomes  $65 \%$ at 40 MeV. The reduction in the case of antineutrino-induced process 
in $^{40}Ar$ is  $60 \%$ at 10 MeV and $56 \%$ at 40 MeV and in case of lead, it is $70 \%$ at 10 MeV and $66 \%$ at 40 MeV. 

 In Fig.(\ref{fig:NC_compare}), we have also presented the results for the total scattering cross-section 
in $^{208}Pb$ for the neutral current neutrino induced process and compared them with the results obtained by 
 Engel {\it et al.}\cite{Engel:2002hg} (dashed line), Kolbe {\it et al.}~\cite{Kolbe:2000np} (dotted line) and 
 Jachowicz {\it et al.}~\cite{Jachowicz:2002hz}(dashed-dotted line). 

 We find that our results for the total cross-sections in the various nuclei like  $^{12}C$, $^{16}O$, $^{40}Ar$ and $^{56}Fe$ are in fair agreement 
 with the other calculations except in the case of $^{208}Pb$ where the results for the charged current induced reactions are lower
 than the results of Engel{\it et al.}~\cite{Engel:2002hg} and Paar {\it et al.}~\cite{Paar:2007fi}. In the case of neutral current induced reactions, our results are 
 higher than the results of Jachowicz {\it et al.}~\cite{Jachowicz:1998fn}, but are in reasonable agreement with the results of 
 Kolbe {\it et al.}~\cite{Kolbe:1992xu} and Engel {\it et al.}~\cite{Engel:2002hg} within 10-15$\%$ for the energy range of 20-60 MeV. 
 The results for the antineutrino-nucleus cross-sections are qualitatively similar to the results of neutrino-nucleus cross-section and 
 within 15-20$\%$ of the results of Jachowicz {\it et al.}~\cite{Jachowicz:1998fn} and Kolbe {\it et al.}~\cite{Kolbe:1992xu}.
 
 Our approach of calculating inclusive cross-sections in the local density approximation is similar to the closure approximation where all the excitations to the 
 final states are summed over. The information about the nuclear structure is simulated through the nuclear density of the initial state
  described by the density parameters fixed by the electron scattering data. The effect of nucleon correlations and meson exchange currents (MEC) are included 
  through RPA using a nucleon-nucleon potential, with $\pi$ and $\rho$ exchanges along with a phenomenological Landau-Migdal parameter to account 
  for the short range correlations. This is an approximate and simple way to describe the nuclear
  medium effects and reproduces correctly the quenching of nuclear response function in the spin-isospin channel. This method has been quite successfully 
  applied to describe the inclusive processes induced by the photons, electrons and muon capture in medium and heavy nuclei. On the other hand, the other 
  papers cited in this work use some explicit model of the nucleon-nucleon interaction to construct the initial and final states to calculate the various transitions
  to the ground state and the higher excited states induced by the Fermi transition, Gamow-Teller(GT) transition as well as to the higher forbidden transitions. 
 The dominant GT transitions are calculated using a shell model or HFB wave functions for ground state and low lying excited states 
  using nucleon-nucleon potential. The strength of the effective weak magnetic and axial couplings are obtained using the phenomenological values of quenching 
  determined experimentally 
  from the electron scattering in the case of vector transitions using CVC and the low energy experimental data on the p-n reactions in the case of axial vector transitions.
   The transitions to the higher forbidden states are calculated by the various authors using different variants of RPA and with the different 
  nucleon-nucleon potentials. The use of the different potentials as well as the different models of RPA brings uncertainty in the prediction of the total cross-sections 
  which is generally about 20$\%$, but could be as large as a factor of 2 in some cases as discussed recently by Balasi {\it et al.}~\cite{Balasi:2015dba}
  and Paar {\it et al.}~\cite{Paar:2007fi,Paar:2012dj}.
  In a nucleus like $^{208}Pb$, there is an additional uncertainty arising due to the treatment of Coulomb effect which could be large specially in the low 
  energy region of $E_\nu < 50$ MeV~\cite{Paar:2007fi}. In view of the above scenario regarding the uncertainty in the theoretical results due to  
  the nuclear medium effects, our results 
  in a simple model presented in Figs.~6-8 for the various nuclei could be considered in agreement with the calculations using explicit nuclear wave functions.
  
  We have also calculated the total scattering cross-section for the neutral current induced 
 processes using the expression of $J^{\mu}$ given in Eq.(\ref{NC_curr}) with the form factors defined in Eq.(\ref{ncff1}) with $F_{1,2}^{s}(q^2)$=0 and 
 $F_A^s(0)$=-0.12 in Eq.(\ref{str-par2}), as well as with the nonstandard interaction using Eq.~(\ref{nsi-expr})
 without the presence of the strangeness form factors, and presented the results in Fig.(\ref{fig:NC_compare1}).
 We find that the effect of including the strangeness form factors, the cross-section for the protons in $^{208}_{82}Pb$ increases by about 20$\%$ 
 and the cross-section for the neutrons in $^{208}_{82}Pb$ decreases by about 10$\%$ from the 
 results without the strangeness contribution but the overall effect in the nuclear targets (protons and neutrons taken together) is very small ($< 1\%$).
The inclusion of NSI interaction results in a significant change in the cross-section. For example, 
 with $\epsilon_{\mu e}^{u V}=\epsilon_{\mu e}^{d V}=$0.05 in Eq. \ref{nsi-expr}, the results of 
   the cross-section for the protons in $^{208}_{82}Pb$ increases by $10\%$ 
 and decreases by 45$\%$ for the neutrons in $^{208}_{82}Pb$, and the overall effect in $^{208}_{82}Pb$ is a decrease in the cross-section by about 28$\%$.
Quantitatively, the contribution of the NSI to the total cross-section is sensitive to the numerical values used for $\epsilon_{\mu e}^{u,d V}$ and would be small 
 for medium nuclei with smaller neutron excess.
 \end{enumerate}
\vspace{-3 mm}
\subsection{Flux-averaged cross-section and event rates}\label{subb}
The flux-averaged cross-section for the Michel spectrum (Eq.(\ref{michel})) and the pion decay in flight $\nu_\mu$ spectrum~\cite{Albert:1994xs} (Eq.(\ref{xsection_final}))
 have been calculated and the results are shown in Fig.(\ref{fig:cc_100_log}) for $^{12}C$ and $^{56}Fe$. In Table \ref{flux_averaged_xsection},
we show our results for the flux-averaged cross-sections and compare them with the experimental results where the available and some theoretical results 
reported in the literature~\cite{Bodmann:1994py,Kolbe:1995af,Athanassopoulos:1997rm,Auerbach:1997ay,Krakauer:1991rf,Nieves:2004wx,Nieves:2017lij,
SajjadAthar:2005ke,Kolbe:2000np,Albert:1994xs,Athanassopoulos:1997rn,Auerbach:2002iy,Samana:2008pt}.
We see that the present model of including various nuclear medium effects compares well with the other theoretical calculations and reproduces the 
experimental results satisfactorily.

The success of our model exhibited in Table \ref{flux_averaged_xsection} in reproducing the flux-averaged cross-sections for the low energy neutrinos has
encouraged us to apply it to the supernova neutrinos/antineutrinos. In Table \ref{event_rates_nue_nuebar}, we show our results for the event rates using flux-averaged 
cross-sections for supernova neutrinos/antineutrinos flux spectra given by Totani {\it et al.}~\cite{Livermore}, Duan {\it et al.}~\cite{FLUX_WEB} 
and Gava {\it et al.}~\cite{Kneller} using Eq.(\ref{fluxavg}).

We see that there is a remarkable variation in the number of events due to the large variation in the three numerically simulated supernova 
$\nu_{e}(\bar\nu_e)$ fluxes considered here. For example, in the case of $\nu_{e}$-induced process, event rate increases when we use the flux of  Gava 
{\it et al.}~\cite{Kneller} instead of Duan {\it et al.}~\cite{FLUX_WEB}. However, there is further increment in the event rates when one uses the flux 
given by Totani {\it et al.}~\cite{Livermore} as compared to Duan {\it et al.}~\cite{FLUX_WEB}. 
 We find similar variation in the event rates for $\bar\nu_e$ induced reactions in these nuclei. 

\subsection{Flux-averaged angular distribution of the outgoing charged lepton}\label{subc}
                                                                  
In Fig.(\ref{fig:cc_dsdcost}), we have presented the results for the flux-averaged angular distribution of the outgoing charged lepton$\left(\left\langle\frac
{d\sigma}{dcos\theta_{\nu l}}\right\rangle\right)$, defined as
\begin{equation}\label{fluxavg_dcost}
\left\langle\frac{d\sigma}{dcos\theta_{\nu l}}\right\rangle = \int \frac{d\sigma}{dcos\theta_{\nu l}} \mathit{f}(E_{\nu}) dE_{\nu},
\end{equation}
where $\frac{d\sigma}{dcos\theta_{\nu l}}$ is
obtained with RPA effect and $\mathit{f}(E_{\nu})$ is the supernova neutrino/antineutrino flux obtained from Refs.~\cite{Livermore}, \cite{FLUX_WEB} and \cite{Kneller}.

In the case of $^{12}C$ and $^{16}O$ nuclear targets, we find that for the flux given by 
Gava {\it et al.}~\cite{Kneller}, the angular distribution is larger than the distribution obtained using the flux of 
Totani {\it et al.}~\cite{Livermore}, and there is remarkable variation in the angular distribution if one uses the flux of Duan {\it et al.} 
\cite{FLUX_WEB}.

In the case of heavier mass target like $^{40}Ar$, $^{56}Fe$ and $^{208}Pb$, angular distribution obtained by using the flux given by Gava {\it et al.}
\cite{Kneller} 
is higher by a factor of around $10-15$ than the distribution obtained by using the flux of Duan {\it et al.} \cite{FLUX_WEB}. While the angular 
distribution obtained by using 
the flux of Totani {\it et al.} \cite{Livermore} is higher by a factor of around $15-20$ as compared to the angular
distribution obtained by using the flux of Duan {\it et al.} \cite{FLUX_WEB}. 

Similar are the observations for $\bar\nu_e$ induced processes.

\subsection{Flux-averaged energy distribution of the outgoing charged lepton}\label{subd} 

In Fig.(\ref{fig:cc_dsdel}), we have presented the results for the flux-averaged energy distribution of the outgoing charged lepton$\left(\left\langle
\frac{d\sigma}{dE_{l}}\right\rangle\right)$, defined as
\begin{equation}\label{fluxavg_del}
\left\langle\frac{d\sigma}{dE_{l}}\right\rangle =\int \frac{d\sigma}{dE_{l}} \mathit{f}(E_{\nu}) dE_{\nu},
\end{equation}
where $\frac{d\sigma}{dE_{l}}$ is 
obtained with RPA effect and $\mathit{f}(E_{\nu})$ is the supernova neutrino flux obtained from Refs.\cite{Livermore, FLUX_WEB, Kneller}. 

The results obtained using the different supernova $\nu_e$ and $\bar \nu_e$ fluxes are compared with each other. We observe that the lepton energy 
distribution (Fig.(\ref{fig:cc_dsdel})) obtained by using the flux given by Gava {\it et al.} \cite{Kneller} is much larger than the 
distribution obtained by using the flux given by Totani {\it et al.} \cite{Livermore} with an energy shift in the peak
region in the case of neutrino-induced process. It may also be observed that the distribution calculated using the flux given by Duan {\it et al.} 
\cite{FLUX_WEB} is small as compared to the distribution obtained by using the flux of Gava {\it et al.} \cite{Kneller}
and Totani {\it et al.} \cite{Livermore}.

However, in the case for $\bar\nu_e$ induced process, we observe that the lepton energy distribution (Fig.(\ref{fig:cc_dsdel})) obtained by using the 
flux given by Totani {\it et al.} \cite{Livermore} is much larger than the 
distribution obtained by using the flux given by Gava {\it et al.} \cite{Kneller}.


\section{Summary and Conclusions}
In this work, we have studied inclusive charged current and neutral current induced reactions for supernova neutrino/antineutrino in nuclei like 
$^{12}C$, $^{16}O$, $^{40}Ar$, $^{56}Fe$ and $^{208}Pb$. 
The calculations are done using local Fermi gas model which takes into account nuclear medium effects due to Pauli blocking, Fermi motion
 as well as the renormalization of weak transition strengths in the nuclear medium. The effect of Coulomb 
distortion of the outgoing charged lepton produced in the charged current reactions is taken into account by using the modified effective momentum 
approximation (MEMA) \cite{Engel:1997fy}.
The model is shown to explain successfully the experimentally observed low energy neutrino-nucleus cross-sections for $^{12}C$ and $^{56}Fe$ 
in the case of neutrinos obtained
from muons decay at rest(MDAR) and pions decay in flight ($\pi$DIF) at KARMEN~\cite{Bodmann:1994py}, LSND~\cite{Athanassopoulos:1997rm} and 
LAMPF~\cite{Krakauer:1991rf}. It is therefore quite suitable method to study the low energy $\nu_e$ and $\bar\nu_e$ 
reactions in nuclei relevant for the supernova neutrino/antineutrino energies.

We have presented the numerical results for total cross-sections for neutrino/antineutrino induced charged current and neutral current processes in 
$^{12}C$, $^{16}O$, $^{40}Ar$, $^{56}Fe$ and $^{208}Pb$ at low energies relevant for supernova $\nu_e$ and $\bar\nu_e$. Using these cross-sections, the 
event rates are obtained for the charged lepton production for 
 the theoretical simulations available in the literature for supernova $\nu_e$ and $\bar\nu_e$ fluxes given by Totani {\it et al.} \cite{Livermore}, 
 Duan {\it et al.} \cite{FLUX_WEB} and Gava {\it et al.} \cite{Kneller}.
 We have also calculated flux-averaged cross-section using the spectrum of the neutrino/antineutrino
for these spectra and the charged lepton event/sec corresponding to 1 kT of target material. The numerical results for the angular distribution and 
energy distribution of the outgoing charged lepton produced in these reactions are also presented.

We conclude that
\begin{enumerate}[label=(\roman*)]
 \item The nuclear medium effects like Pauli blocking and Fermi motion effects lead to substantial reduction in the cross-section as compared to the free 
nucleon cross-section. The energy dependence 
of the reduction of the cross-section due to nuclear medium effects is quantitatively different for neutrino and antineutrino induced processes. 
 The $Q$-values and the Coulomb effect of charged lepton play an important role in quantitatively predicting the cross-sections in  $^{12}C$, $^{16}O$, 
 $^{40}Ar$, $^{56}Fe$ and $^{208}Pb$ in the region of low energies relevant for the supernova $\nu_e$ and $\bar\nu_e$. Our results of nuclear medium 
 effect on $\nu$($\bar\nu$)-nucleus cross-sections are in qualitative agreement with the other calculations.
 \item Large variations in the predicted fluxes of supernova neutrino/antineutrino as obtained by the simulation analysis of Totani {\it et al.} 
\cite{Livermore}, Duan {\it et al.} \cite{FLUX_WEB} and 
Gava {\it et al.} \cite{Kneller} lead to large variations in the various observables like event rates, angular and energy distributions of the charged 
leptons when averaged over the neutrino/antineutrino
fluxes. A quantitative description of these observables has been presented for the case of $^{12}C$, $^{16}O$, $^{40}Ar$, $^{56}Fe$ and $^{208}Pb$
nuclei proposed to be used as detector material in future supernova neutrino detectors.
\end{enumerate}

\section{Acknowledgments}
S. Chauhan is thankful to University Grant Commission for providing financial assistance under UGC - Start up grant(No. F.30-90/2015(BSR)).
\section*{Appendices}
\subsection*{Appendix A}
\subsubsection{Isovector vector and axial vector nucleon form factors}
The isovector vector form factors $F_{1,2}^V(q^2)$ are written in terms of the  
electromagnetic Dirac and Pauli form factors of proton (neutron), i.e., $F_{1}^{p(n)}(q^2)$ and $F_{2}^{p(n)}(q^2)$, respectively, as
\begin{equation}\label{f1v_f2v}
F_{1,2}^V(q^2)=F_{1,2}^p(q^2)- F_{1,2}^n(q^2).
\end{equation}
The Dirac and Pauli
form factors are expressed in terms of the experimentally 
determined Sach's electric $G_E^{p(n)}(q^2)$ and magnetic $G_M^{p(n)}(q^2)$ form factors defined as \cite{Stoler:1993yk}
\begin{eqnarray}\label{f1pn_f2pn}
F_1^{p(n)}(q^2)&=&\left(1-\frac{q^2}{4M^2}\right)^{-1}~\left[G_E^{p(n)}(q^2)-\frac{q^2}{4M^2}~G_M^{p(n)}(q^2)\right],\\
F_2^{p(n)}(q^2)&=&\left(1-\frac{q^2}{4M^2}\right)^{-1}~\left[G_M^{p(n)}(q^2)-G_E^{p(n)}(q^2)\right].
\end{eqnarray}
The electric and magnetic Sach's form factors i.e. $G_E^{p(n)}(q^2)$ and $G_M^{p(n)}(q^2)$ are taken 
from the different parameterizations available in the literature~\cite{Galster:1971kv, Bradford:2006yz, Budd:2005tm, Bosted:1994tm, Alberico:2008sz}. For example, in the 
parameterization given by Galster {\it et al.}~\cite{Galster:1971kv}, Sach's form factors are defined as
\begin{eqnarray}
G_E^p(q^2)&=&\frac{1}{(1-q^2/M_v^2)^2}, \nonumber \\ 
G_M^p(q^2)&=&(1+\mu_p)G_E^p(q^2), \nonumber\\
G_M^n(q^2)&=&\mu_nG_E^p(q^2), \nonumber\\
G_E^n(q^2)&=&(\frac{q^2}{4M^2})\mu_nG_E^p(q^2)\xi_n, \nonumber\\
\xi_n &=& \frac{1}{(1-\lambda_n\frac{q^2}{4M^2})}, \nonumber
\end{eqnarray}
where $ \mu_p=1.7927\mu_N,~      \mu_n=-1.913\mu_N,~        M_v=0.84GeV, ~    \lambda_n=5.6,~ \mu_N$ is the nucleon magnetic moment. 
$\mu_p~\rm{and}~\mu_n$ stand for the proton and neutron anomalous magnetic moment, respectively. 

The isovector axial form factor $F_A^V(q^2)$ is parameterized as
\begin{equation}\label{fa}
F_A^V(q^2)=F_A(0)~\left[1-\frac{q^2}{M_A^2}\right]^{-2},
\end{equation}
where $F_A(0)$ is the axial charge and $M_A$ is the axial dipole mass. For the numerical calculations, we have taken $F_A(0)$ = - 1.267 and $M_A$ = 
1.05 GeV.

The pseudoscalar form factor $F_p^V(q^2)$ is given in terms of $F_A^V(q^2)$ using the Goldberger-Treiman relation as 
\begin{equation}\label{fp}
F_p^V(q^2)=\frac{2MF_A^V(q^2)}{m_\pi^2-q^2},
\end{equation}
where $m_\pi$ is the mass of pion.
\subsubsection{Vector and axial vector form factors with strangeness of the nucleon}
 $\tilde F_{1,2}^{N}(q^2)$ are defined in terms of the standard electromagnetic Dirac and Pauli form factors of the nucleon $F_{1}^{p,n}(q^2)$ 
and $F_{2}^{p,n}(q^2)$ and 
a strange vector component $F_{1,2}^{s}(q^2)$, in the following way \cite{Alberico:1997rm}:
\begin{equation}\label{ncff1}
\left.
 \begin{array}{c}
  \tilde F_{1,2}^{p}(q^2) = (\frac{1}{2} - 2 sin^{2}\theta_{W})F_{1,2}^{p}(q^2) - \frac{1}{2} F_{1,2}^{n}(q^2) - \frac{1}{2} F_{1,2}^{s}(q^2),\\
  \\
  \tilde F_{1,2}^{n}(q^2) = (\frac{1}{2} - 2 sin^{2}\theta_{W})F_{1,2}^{n}(q^2) - \frac{1}{2} F_{1,2}^{p}(q^2) - \frac{1}{2} F_{1,2}^{s}(q^2)
 \end{array}
\right\}
\end{equation}
 
 and the axial form factors $\tilde F_{A}^{p,n}(q^2)$ are given by

 \begin{equation}\label{fanc}
\tilde F_{A}^{p,n}(q^2) = \pm \frac{1}{2} F_{A}(q^2) + \frac{1}{2} F_{A}^{s}(q^2),
\end{equation}
 where $\theta_{W}$ is the weak mixing angle. For the numerical calculations, we have taken $sin^{2}\theta_{W}$ = 0.2315~\cite{Olive:2016xmw}.
 + sign is for the proton target and - sign is for the neutron target,
 $F_A(q^2)$ is given by Eq.(\ref{fa}). $F_{A}^{s}(q^2)$ is the strange axial vector form factor.
 
 The vector and axial vector strangeness form factors $F_{1s} (q^2)$, $F_{2s} (q^2)$ and $F_{A}^s (q^2)$ are parameterized as~\cite{Maas:2017snj}
 
 \begin{eqnarray}
 \label{str-par}
  F_1^s(q^2) &=& -F_1^s(0) ~q^2~ \frac{1}{1+\tau}~ \frac{1}{(1-q^2/M_v^2)^2} , \\
   \label{str-par1}
  F_2^s (q^2) &=& F_2^s(0)~ \frac{1}{1+\tau} ~\frac{1}{(1-q^2/M_v^2)^2} ,\\
   \label{str-par2}
  F_A^s (q^2) &=& F_A^s (0) ~\frac{1}{(1-q^2/M_A^2)^2}.
 \end{eqnarray}

 The information about the strangeness-dependent quantities viz. $F_1^s(0)$, $F_2^s(0)$ and $F_A^s (0)$ are obtained from the parity violating observables in the 
 scattering of polarized electron scattering as well as from the neutral current-induced (anti)neutrino scattering from the nucleon and the nuclear 
 targets. The experiments with the polarized electrons from the nucleon and the nuclear targets in the kinematic region of elastic, DIS and SIDIS scattering 
 are consistent with $F_1^s(0)$=0 and a small value of $F_2^s(0)$ which is statistically consistent with $F_2^s(0)$=0. The limits on $F_A^s (0)$ range from 
 $- 0.2 $ to $+ 0.08$~\cite{Maas:2017snj}. On the other hand the two experiments with (anti)neutrino beams at 
 MiniBooNE~\cite{AguilarArevalo:2010cx} and BNL~\cite{Garvey:1992cg, Ahrens:1986xe} from nuclear targets assuming $F_1^s(0)$ = $F_2^s(0)$ =0
 give a range of $F_A^s (0)$ varying from $0.08$ to $-0.16$. In 
 view of this uncertainty in the strangeness form factors, we have used $F_1^s(0)$ = $F_2^s(0)$ =0 and $F_A^s (0) = -0.12$ in the numerical calculations.

\subsubsection{Proton and neutron form factors with nonstandard interaction}
  $F_{1,2}^{p}(q^2)$ and $F_{1,2}^{n}(q^2)$  weak nucleon vector form factors with nonstandard interaction assuming $\epsilon_{e \mu}^{qL} = \epsilon_{e \mu}^{qR}$
  are given by ~\cite{Papoulias:2016edm}
 \begin{eqnarray}\label{nsi-expr}
  \tilde F_{1,2}^{p}(q^2) \rightarrow \tilde {F_{1,2}^{p}}^\prime(q^2) = \tilde F_{1,2}^{p}(q^2) + \varepsilon_{\mu e}^{p V}(Q^2) ,\nonumber\\
\tilde F_{1,2}^{n}(q^2) \rightarrow \tilde {F_{1,2}^{n}}^\prime(q^2) = \tilde F_{1,2}^{n}(q^2) - \varepsilon_{\mu e}^{n V}(Q^2),
\end{eqnarray}
where
\begin{eqnarray*}\label{nsi-1}
\varepsilon_{\mu e}^{p V}(Q^2) &=& (2 \epsilon_{\mu e}^{u V} + \epsilon_{\mu e}^{d V}) \left( 1 - \frac{q^2}{M_V^2} \right)^{-2}, \nonumber \\
\varepsilon_{\mu e}^{n V}(Q^2) &=& (\epsilon_{\mu e}^{u V} + 2\epsilon_{\mu e}^{d V}) \left( 1 - \frac{q^2}{M_V^2} \right)^{-2},
\end{eqnarray*}
and $\epsilon_{\mu e}^{u V}=\epsilon_{\mu e}^{d V}=$0.05, $M_V$=0.84GeV, are used for the numerical calculations~\cite{Papoulias:2016edm}.

\subsection*{Appendix B}
\begin{figure}
\includegraphics[height=.3\textheight,width=0.64\textwidth]{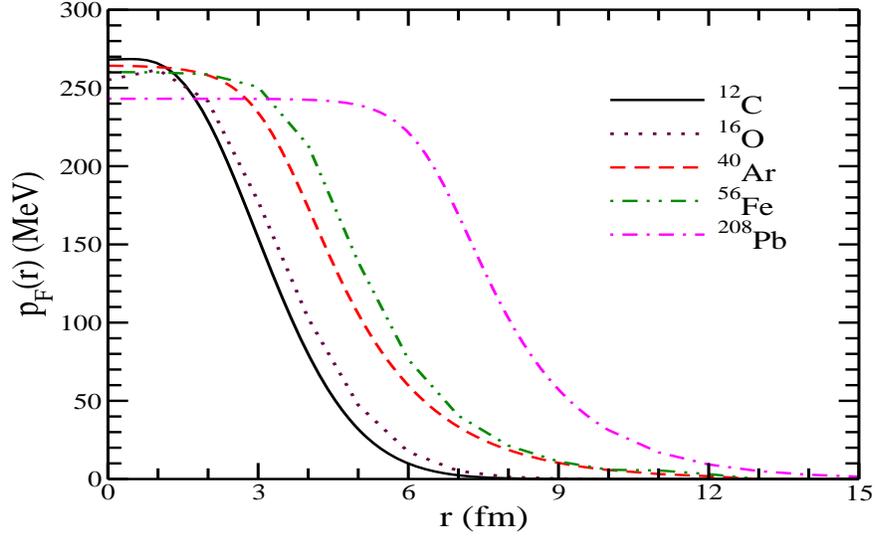}
\caption{Fermi momentum $p_{F}(r)$ versus r for various nuclei.}
\label{fig:Fermi_mom}
\end{figure}

We use the proton density $\rho_p (r) = \frac{Z}{A}\rho (r)$ and neutron density given by $\rho_n (r) = \frac{A-Z}{A}\rho (r)$, where $\rho (r)$ is the 
 nuclear density determined
 experimentally by the electron-nucleus scattering experiments~\cite{Vries:1974, Vries:1987}. We use the modified harmonic oscillator(MHO) density 
 \begin{eqnarray}\label{MHO}
 \rho(r) = \rho(0) \left[ 1 + a \left( \frac{r}{R} \right)^{2} exp\left[ -\left( \frac{r}{R} \right)^{2} \right] \right]
 \end{eqnarray}
for $^{12}C$ and $^{16}O$ and two-parameter Fermi density (2pF) 
\begin{eqnarray}\label{2pF}
 \rho(r) = \frac{\rho(0)} {\left[ 1 +  exp\left( \frac{r-R}{a} \right)\right]}
 \end{eqnarray}
for $^{40}Ar$, $^{56}Fe$ and $^{208}Pb$ with $R$ and $a$ as density parameters. In Table \ref{tab:Q-value_BE}, we show the nuclear density and other parameters needed 
for numerical calculations in this paper. In Fig.(\ref{fig:Fermi_mom}), 
 we have shown Fermi momentum ($p_{F}(r)$) as a function of position ($r$) for various nuclei used in this work.

 \subsection*{Appendix C}
The Lindhard function for the particle-hole excitation corresponding to Fig.(\ref{fig:neutrinoselfenergy}) is given by 
\begin{equation}
U_N(q_{0}^{\nu(\bar\nu)},{\bf q}) = 2\;\int \frac{d{\bf p}}{(2\pi)^3} \frac{M_{n}M_{p}}{E_{n}({\bf p})E_{p}(\bf p + \bf q)} \left(\frac{n_{i}({\bf p})
[1-n_{f}({\bf p} + {\bf q})]}{q_{0}^{\nu(\bar\nu)}+E_{n}({\bf p})-E_{p}(\bf p+\bf q)+i\epsilon }\right),
\end{equation}
where $i(f)=n(p)$ for $\nu$ and $p(n)$ for $\bar\nu$ scattering and $q_0^{\nu(\bar\nu)}=q_0-Q^{\nu(\bar\nu)}$. $Q^{\nu(\bar\nu)}$ being the $Q$ value of 
${\nu(\bar\nu)}$ reactions given in Table \ref{tab:Q-value_BE} for various nuclei. 
$n_i({\bf p})$ is the occupation number of the neutron (proton) and $n_f({\bf p +q })$ is the occupation number of the proton (neutron). 
\begin{table}
\begin{tabular}{|c|c|c|c|c|c|c|}\hline\hline
Nucleus & Binding Energy & Q-Value($\nu$)& Q-Value($\bar\nu$)& $R_p$ & $R_n$ & a \\
        &    (MeV)       &    (MeV)      &    (MeV) & (fm)\cite{Nieves:2004wx}&(fm)\cite{Nieves:2004wx}&$(fm)^\ast$\cite{Nieves:2004wx}\\ \hline
$^{12}C$& 25 & 17.84 & 13.90 &1.69 &1.692  &1.082(MHO)\\
$^{16}O$& 27 & 19.70 & 14.30 & 1.83&1.833 &1.544(MHO) \\
$^{40}Ar$& 30 & 3.64 & 8.05 & 3.47& 3.64 & 0.569(2pF)\\
$^{56}Fe$& 36 & 6.52 & 4.35 & 3.97 & 4.05 & 0.593(2pF)\\
$^{208}Pb$& 44 & 5.20 & 5.54 & 6.62 & 6.89 & 0.549(2pF)\\\hline
\end{tabular}
\caption{Binding energy, Fermi momentum and Q-value of the reaction for various nuclei. Last three columns are the parameters for MHO and 2pF densities.
 $^{\ast}$ is dimensionless for the MHO density.}
 \label{tab:Q-value_BE}
\end{table}

 Using 
\begin{equation}
 {1 \over \omega \pm i \eta} = {\cal P}({1 \over \omega}) \mp i \pi \delta(\omega), \nonumber
\end{equation}
 the imaginary part of the Lindhard function is written as
\begin{eqnarray}
 Im \; U_N(q_{0}^{\nu(\bar\nu)},{\bf q}) &=& 2\;\int \frac{d{\bf p}}{(2\pi)^3}\; n_i({\bf p})[1- n_f({\bf p +q })] \;  \frac{\pi M_{n}M_{p}}{E_{n}({\bf p})E_{p}(\bf p + \bf q)} \nonumber \\
 &\times& \delta^0(q_0^{\nu(\bar\nu)} + \sqrt{|{\bf p}|^2+M^2} - \sqrt{|{\bf p}|^2+|{\bf q}|^2 + 2|{\bf p}| |{\bf q}| cos\theta +M^2})\;,
\end{eqnarray}
which after simplification may be written as~\cite{Singh:1993rg}
\begin{equation}
Im ~ U_N (q_{0}^{\nu(\bar\nu)}, {\bf q}) = -\frac{1}{2\pi} \frac{M_{p} M_{n}}{|{\bf q}|} [E_{F_{1}} - A]
\end{equation}
with
\begin{equation}
 cos\theta = \frac{ (q_0^{\nu/\bar\nu})^2 -|{\bf q}|^2 +2 q_0^{\nu/\bar\nu} \sqrt{|{\bf p}|^2+M^2}}{2 |{\bf p}| |{\bf q}|} \le 1 , \nonumber
\end{equation}
$q^2 < 0, \;  E_{F_2} - q_{0}^{\nu(\bar\nu)} < E_{F_{1}}$ and $\frac{-q_{0}^{\nu(\bar\nu)} + |{\bf q}|\sqrt{1 - \frac{4 M^2}{q^2} }}{2} < E_{F_{1}}$, 
where $E_{F_{1}} = \sqrt{p_{F_{n}}^2 + M_{n}^{2}}$, 
$E_{F_{2}} = \sqrt{p_{F_{p}}^2 + M_{p}^{2}}$ and

\begin{equation}
A = Max \left[ M_{n}, E_{F_{2}} - q_{0}, \frac{-q_{0}^{\nu(\bar\nu)} + |{\bf q}|\sqrt{1 - \frac{4 M^2}{q^2} }}{2}\right].
\end{equation}
Otherwise, $Im ~U_N = 0$.


%
%
%

%
%
%
%

%
%
%
\end{document}